\documentclass{jfm}
\usepackage{graphicx}
\usepackage{epstopdf, epsfig}

\usepackage{multiJFMequations}
\usepackage{amsmath,bm}
\usepackage{color}
\newcommand*{\be}{\begin{equation}}
\newcommand*{\ee}{\end{equation}}
\newcommand*{\bse}{\begin{subequations}}
\newcommand*{\ese}{\end{subequations}}
\newcommand*{\bme}{\begin{multiequations}}
\newcommand*{\eme}{\end{multiequations}}
\newcommand*{\se}{\singleequation}
\newcommand*{\de}{\doubleequation}
\newcommand*{\te}{\tripleequation}

\newcommand*{\XXint}[3]{{\setbox0=\hbox{$#1{#2#3}{\int}$}
\vcenter{\hbox{$#2#3$}}\kern-.5\wd0}}

\newcommand*{\XXnotinfty}[3]{{\setbox0=\hbox{$#1{#2#3}{\to}$}
\vcenter{\hbox{$#2#3$}}\kern-.5\wd0}}

\newcommand*{\ds}{\displaystyle}
\providecommand*{\dfrac}[2]{\ds\frac{#1}{#2}}

\renewcommand*{\Re}{\mbox{Re}}
\renewcommand*{\Im}{\mbox{Im}}

\renewcommand*{\tilde}{\widetilde}
\renewcommand*{\hat}{\widehat}
\renewcommand*{\bar}{\overline}

\newcommand*\bfcdot{\bm{\cdot}}
\newcommand*\bfnabla{\bm{\nabla}}
\renewcommand*{\sp}[2]{{#1}\bfcdot{#2}}
\newcommand*{\vp}[2]{{#1}\times{#2}}
\newcommand*{\od}[2]{\dfrac{{\mathrm d}{#1}}{{\mathrm d}{#2}}}
\newcommand*{\pd}[2]{\dfrac{\partial{#1}}{\partial{#2}}}

\newcommand*{\erfc}{\,{\mbox{erfc}}\,}

\renewcommand*{\Lambda}{\varLambda}
\renewcommand*{\Upsilon}{\varUpsilon}
\renewcommand*{\Phi}{\varPhi}
\renewcommand*{\Psi}{\varPsi}
\renewcommand*{\Omega}{\varOmega}
\renewcommand*{\Theta}{\varTheta}
\renewcommand*{\Xi}{\varXi}
\newcommand*{\Omegav}{{\bm{\Omega}}}
\newcommand*{\dR}{{\mathrm d}}\newcommand*{\eR}{{\mathrm e}}\newcommand*{\iR}{{\mathrm i}}
\newcommand*{\HR}{{\mathrm H}}\newcommand*{\IR}{{\mathrm I}}\newcommand*{\JR}{{\mathrm J}}\newcommand*{\LR}{{\mathrm L}}\newcommand*{\OR}{{\mathrm O}}
\newcommand*{\uv}{{\bm{u}}}\newcommand*{\zv}{{\bm{z}}}

\newcommand*{\tS}{{\sf{t}}}\newcommand*{\zS}{{\sf{z}}}
\newcommand*{\LS}{{\sf{L}}}\newcommand*{\YS}{{\sf{Y}}}\newcommand*{\ZS}{{\sf{Z}}}
\newcommand*{\DC}{{\mathcal D}}

\newcommand*{\Upsilonh}{{\hat \Upsilon}}\newcommand*{\Xih}{{\hat \Xi}}

\newcommand*{\tANS}{{\mbox{\tiny {ANS}}}}
\newcommand*{\tENS}{{\mbox{\tiny {ENS}}}}
\newcommand*{\tDNS}{{\mbox{\tiny {DNS}}}}
\newcommand*{\tFS}{{\mbox{\tiny {FS}}}}

\shorttitle{A spin-down problem}
\shortauthor{L. Oruba, A.~M. Soward and E. Dormy}

\title{Spin-down in a rapidly rotating cylinder container with mixed rigid and stress-free boundary conditions}

\author{L. Oruba \aff{1}
  A. M. Soward \aff{2}
  \corresp{\email{andrew.soward@ncl.ac.uk}},
 \and E. Dormy\aff{3}}

\affiliation{\aff{1}D\'{e}partement de Physique, \'{E}cole Normale Sup\'{e}rieure, 24 rue Lhomond, 75005 Paris, FRANCE
\aff{2} School of Mathematics and Statistics, Newcastle University, Newcastle upon Tyne NE1 7RU, UK
\aff{3} {D\'epartement de Math\'ematiques et Applications, UMR-8553, \'Ecole Normale Sup\'erieure, 45 rue d'Ulm, 75005 Paris, FRANCE}}

\begin{document}

\maketitle

\begin{abstract}
A comprehensive study of the classical linear spin-down of a constant density viscous fluid (kinematic viscosity $\nu$) rotating rapidly (angular velocity $\Omega$) inside an axisymmetric cylindrical container (radius $L$, height $H$) with rigid boundaries, that follows the instantaneous small change in the boundary angular velocity at small Ekman number $E=\nu/H^2\Omega\ll 1$, was provided by Greenspan \& Howard ({\itshape{J.~Fluid Mech.}}, vol.~17, 1963, pp.~385--404). $E^{1/2}$--Ekman layers form quickly triggering inertial waves together with the dominant spin-down of the quasi-geostrophic (QG) interior flow on the $O(E^{-1/2}\Omega^{-1})$ time-scale. On the longer lateral viscous diffusion time-scale $O(L^2/\nu)$, the QG-flow responds to the $E^{1/3}$--side-wall shear-layers. In our variant the side-wall and top boundaries are stress-free; a setup motivated by the study of isolated atmospheric structures, such as tropical cyclones, or tornadoes. Relative to the unbounded plane layer case, spin-down is reduced (enhanced) by the presence of a slippery (rigid) side-wall. This is evinced by the QG-angular velocity, $\omega^\star$, evolution on the $O(L^2/\nu)$ time-scale: Spatially, $\omega^\star$ increases (decreases) outwards from the axis for a slippery (rigid) side-wall; temporally, the long-time$\,(\gg L^2/\nu)$ behaviour is dominated by an eigensolution with a decay rate slightly slower (faster) than that for an unbounded layer. In our slippery side-wall case, the $E^{1/2} \times E^{1/2}$ corner region that forms at the side-wall intersection with the rigid base is responsible for a $\ln E$ singularity within the $E^{1/3}$--layer causing our asymptotics to apply only at values of $E$ far smaller than can be reached by our Direct Numerical Simulation (DNS) of the entire spin-down process. Instead, we solve the $E^{1/3}$--boundary-layer equations for given $E$ numerically. Our hybrid asymptotic-numerical approach yields results in excellent agreement with our DNS.
\end{abstract}

\section{Introduction}

Intense nearly axisymmetric vortices often develop in geophysical flows: tornadoes, or hurricanes in the atmosphere, while in the ocean, the Sea Surface Height variability appears dominated by westward-propagating mesoscale eddies throughout most of the World Ocean \citep{CSS11}. In order to understand their characteristics, it is instructive to model such objects as isolated structures \citep[see][and references therein]{PMSM15}. Their natural embedding inside a cylindrical domain introduces the need for an artificial outer circular side-wall boundary. Though no-slip boundaries could be contemplated, they introduce additional friction and so nowadays other boundary conditions are usually considered. Whatever the model, boundary layers are important \citep[see, e.g.,][]{SM10}.

\cite{W68} was the first to suggest removing some frictional constraints associated with the side-walls in the numerical models, in order to more faithfully mimic the unbounded physical domain. Viscous friction was then only retained at the bottom boundary, as in atmospheric flows. Since then, stress-free side-walls, at which the angular velocity gradient vanishes, have been used in several numerical studies \citep[see, e.g.,][]{R86a,R86b}; while matching to an external azimuthal flow corresponding to an axial line vortex (no viscous force), is sometimes preferred \citep[see, e.g.,][]{MSY01}.
  
Here, we investigate the effects of a stress-free side-wall on such flows. Specifically, we consider the classical linear spin-down problem (see, e.g., \cite{GH63} and the review of \cite{BC74}; for non-linear studies see, e.g., \cite{W64,HLFW83} and the review of \cite{DF01}) in a domain with no-slip boundary at the bottom but modified by the presence of stress-free side-wall and top boundaries. Like \cite{GH63}, we find that during the early stages, the mainstream spin-down exterior to all boundary layers is characterised by an angular velocity that is spatially constant but decays exponentially in time just as in an infinite (unbounded) plane layer (see \S\ref{GH63-problem}). However on a somewhat longer time-scale the quasi-geostrophic angular velocity develops a radial (non-constant) structure dependent on the jump condition across the \cite{S57} $E^{1/3}$--side-wall shear-layer. In the rigid boundary case studied by \cite{GH63}, that shear-layer is passive and the quasi-geostrophic angular velocity simply vanishes at the mainstream boundary. This may be interpreted as a retrograde torque at the outer boundary  that enhances the decay of the angular velocity with increasing radius. In our stress-free boundary case, the torque at the true boundary certainly vanishes. Nevertheless, the return meridional flow in the  $E^{1/3}$-layer is necessarily opposite to that needed to achieve spin-down in the mainstream. So from that point of view it is perhaps not surprising to find that the outer boundary essentially opposes spin-down rather than enhancing it. Essentially the $E^{1/3}$-layer is no longer passive but completely alters the effective mainstream boundary condition. Consequently, the mainstream flow experiences a prograde boundary-torque proportional to the quasi-geostrophic angular velocity there. In turn, that causes the angular velocity to increase (rather than decrease) outwards (see (\ref{ms-bc})). This was an unexpected finding; though with hindsight it is easily explained.

\subsection{The Greenspan and Howard problem\label{GH63-problem}}

Here we describe in more detail our variant of the \cite{GH63} model. Relative to cylindrical polar coordinates, $(r^\star,\,\theta^\star,\,z^\star)$, we consider a cylindrical container of height $H$ and radius $L$ rotating rapidly with angular velocity $\Omegav$ about its axis of symmetry. The container is filled with constant density fluid of viscosity $\nu$, which initially at time $t^\star=0$ rotates rigidly with angular velocity $(1+\varepsilon)\Omegav$, where $0<\varepsilon\ll 1$. The top boundary ($r^\star<L$, $z^\star=H$) and the side-wall ($r^\star=L$, $0<z^\star<H$) are impermeable and stress-free. The lower boundary ($r^\star<L$, $z^\star=0$) is rigid. For that reason alone the initial state of rigid rotation  $(1+\varepsilon)\Omegav$ of the fluid cannot persist and the fluid spins down to the final state of rigid rotation $\Omegav$ of the container as $t^\star\to\infty$. 

The rapid rotation of the system is measured by the small Ekman number
\be
\label{Ek-numb}
E\,=\,\nu \big/\bigl(H^2\Omega\bigr)\,\ll\, 1\,.
\ee
On very short time scales shear-layers form adjacent to the all boundaries of width
\be
\label{diff-width}
\delta^\star(t^\star)\,=\,(\nu t^\star)^{1/2}\,=\,H(E\Omega t^\star)^{1/2}
\ee
due to viscous diffusion. On the short inertial wave time-scale $t^*=1/\Omega$, a quasi-steady Ekman layer of width $\delta^\star_E=(\nu/\Omega)^{1/2}=E^{1/2}H$ forms adjacent to the lower boundary $z^\star=0$. Various inertial waves and quasi-geostrophic ($z^\star$-independent; QG) motions exterior to all boundary layers are generated in the mainstream which decay primarily due to the ensuing Ekman suction into (or blowing out of) the Ekman layer. Our numerical results generally show that motion is soon dominated by the QG-flow which spins down on the longer time-scale $H\big/(\nu\Omega)^{1/2}=E^{-1/2}/\Omega$. Indeed, relative to the frame rotating with angular velocity $\Omegav$, the cylindrical components of the mainstream velocity are simply 
\bme
\label{simple-spin-down}
\be
\se
\bigl(\,\bar{{\breve u}^\star}\,,\,\bar{{\breve v}^\star}\,,\,\bar{{\breve w}^\star}\,\bigr)=\,\varepsilon \Omega\kappa\Bigl(\tfrac12 \sigma E^{1/2}r^\star,\,r^\star\,,\sigma E^{1/2}\bigl(H-z^\star\bigr)\Bigl)\exp\bigl(-\,\sigma E^{1/2}\Omega t^\star\bigr),
\ee
as determined by the solution to the initial value problem for a layer of unbounded horizontal extent outlined in appendix~\ref{infinite-layer} and previously considered by \cite{GH63}. The purpose of appendix~\ref{infinite-layer} is to extend Greenspan \& Howards' $\kappa\approx 1$, $\sigma\approx 1$ results to determine their more precise forms (\ref{inf-layer-sol}$b$) and (\ref{p-sigma}$a$) respectively. They have expansions
\be
\de
\kappa\,=\,1\,+\,\tfrac14 E^{1/2}\,+\,O(E)\,, \qquad\qquad
\sigma\,=\,1\,+\,\tfrac34 E^{1/2}\,+\,O(E)
\ee
\eme
(see (\ref{kappa-mu}$a$) and (\ref{p-sigma}$b$) respectively), which prove useful when comparing our numerical results at finite $E$ with our asymptotic predictions. The spin-down motion (\ref{simple-spin-down}$a$) is characterised by a fluid flux inside the Ekman layer towards the axis which is blown out into the mainstream with velocity $\bar{{\breve w}^\star}\big|_{z^\star=0}=\epsilon\Omega \kappa\sigma E^{1/2}H$. This generates a radial outflow in the mainstream of magnitude  $\bar{{\breve u}^\star}\big|_{z^\star>0}=\tfrac12 {\epsilon\Omega}\kappa\sigma E^{1/2}Hr^\star$, which by conservation of angular momentum causes the angular velocity decay
\be
\label{omegabar}
\bar{{\breve \omega^\star}}(t^\star)\,=\,\bar{{\breve v^\star}}/r^\star\,=\,\varepsilon \Omega\kappa\exp\bigl(-\,\sigma E^{1/2}\Omega t^\star\bigr).
\ee

The mainstream outflow is blocked by the outer boundary $r^\star=L$, where a quasi-steady \cite{S57} side-wall layer of width $\delta^\star_S=E^{1/3}H$ forms \citep[see][]{B68} on the time-scale $t^*=E^{-1/3}/\Omega$, long compared to the Ekman layer time-scale $1/\Omega$ but short compared to the spin-down time $E^{-1/2}/\Omega$. Following the formation of the $E^{1/3}$-Stewartson layer, an ever thickening QG-shear-layer emerges of width $\delta^\star(t^\star)=H(E\Omega t^\star)^{1/2}$ (see (\ref{diff-width})). On the spin-down time-scale $E^{-1/2}/\Omega$, it has width $E^{1/4}H$, which prompted \cite{GH63} to refer to it as an $E^{1/4}$-layer. The implied link with the static $E^{1/4}$-Stewartson layer is misleading as the dynamical balances are different and there are no persistent (quasi-static) $E^{1/4}$-layers in the spin-down problem (see also the discussions below (\ref{QG-eq-b}) and in \S\ref{Discussion}).

Following the establishment of the $E^{1/3}$-layer (i.e., $E^{1/3}\Omega t^\star\gg  1$), the QG-shear-layer thickens: $\delta^\star(t^\star)=H(E\Omega t^\star)^{1/2}$. This modifies the spin-down profile (\ref{omegabar}) both spatially and temporally on the longer lateral diffusion time $L^2/\nu=\ell^2 E^{-1}/\Omega$, where
\be
\label{l-def}
 \ell\,=\,L/H
\ee
is the container aspect ratio. This suggests a two time-scale approach \citep[see, e.g.,][Chapter 7]{H91} in which (\ref{omegabar}) is modulated by a factor $\bar{\mathring \omega}\bigl(r^\star/H,E\Omega t^\star\bigr)$ so taking the form
\be
\label{Ho-Gr}
\bar{{\breve \omega^\star}}(r^\star,t^\star)\,=\,\varepsilon \Omega \kappa\,\bar{\mathring \omega}\bigl(r^\star/H,E\Omega t^\star\bigr)\exp\bigl(-\,\sigma E^{1/2}\Omega t^\star\bigr).
\ee
Matching with the short time solution (\ref{omegabar}), as $E\Omega t^\star\downarrow 0$, is achieved by demanding that initially
\be
\label{ms-ic}
\bar{\mathring \omega}(r^\star/H,0)=1\,.
\ee
The outer $r^\star=L$ boundary condition on the QG-mainstream flow depends on the jump conditions across the $E^{1/3}$--side-wall layer.

For the rigid outer boundary considered by \cite{GH63} the jump in azimuthal velocity across the $E^{1/3}$-layer is negligible and so the outer boundary condition on the mainstream angular velocity is simply $\bar{{\breve \omega^\star}}\bigl(L,t^\star\bigr)=0$. It causes the early rigid rotation profile (\ref{omegabar}) to be further eroded to zero through lateral diffusion inwards from the outer boundary. For our stress-free outer boundary, consideration of the $E^{1/3}$-layer shows that the boundary condition on the mainstream angular velocity is different of mixed type taking the Robin boundary condition form
\be
\label{ms-bc}
\pd{\bar{{\breve \omega}^\star}}{r^\star}\,=\,\alpha \,\dfrac{\bar{{\breve \omega}^\star}}{H}  \qquad\qquad\mbox{at} \qquad r^\star=\,L\,,
\ee
where $\alpha$ depends on the nature of the $E^{1/3}$-layer and is a function of $E$ alone, and $2\pi H L^3(\rho\nu)\partial{\bar{{\breve \omega}^\star}}\big/\partial r^\star\big|_{r^\star=L}$ (viscosity $\rho\nu$) is the total outer (mainstream) boundary torque on the QG-mainstream flow (see also (\ref{QG-bc})). The long time behaviour of the solution has the functional form 
\be
\label{ms-lt-1}
\bar{\mathring \omega}\,\approx\,{\bar\omega}\bigl(r^\star/H\bigr)\exp\bigl(k^2\ell^{-2}E\Omega t^\star\bigr),
\ee
as $\ell^{-2}E\Omega t^\star\to\infty$. Unlike $\alpha$ in (\ref{ms-bc}), the constant $k$ is a function of both $E$ (via $\alpha$) and $\ell$.

The main point that we wish to stress is that whereas (\ref{simple-spin-down}) defines the mainstream eigensolution on the spin-down time $E^{-1/2}/\Omega$ with growth rate $-\,\sigma E^{1/2}\Omega$, that solution evolves over the lateral diffusion time-scale $\ell^2 E^{-1}/\Omega$ and asymptotes to the form 
\bme
\label{ms-lt}
\be
\bar{{\breve \omega}^\star}\,\approx\,\varepsilon \Omega\kappa \,{\bar\omega}\bigl(r^\star/H\bigr)\exp\bigl(-\,qE\Omega t^\star\bigr), \qquad\qquad   q\,=\,\sigma E^{-1/2}-\,k^2\ell^{-2},
\ee
\eme
with the different growth rate $-\,qE\Omega$ and non-constant radial profile proportional to ${\bar\omega}\bigl(r^\star/H\bigr)$. Remarkably the stress-free outer boundary actually hinders the spin-down as evinced by the positive growth rate $k^2\ell^{-2}E\Omega$ of $\bar{\mathring \omega}$ (see (\ref{ms-lt-1})). The phenomenon can be traced to the mainstream boundary torque proportional to $\partial{\bar{{\breve \omega}^\star}}\big/\partial r^\star\big|_{r^\star=L}$ determined by (\ref{ms-bc}) with $\alpha>0$. This long time behaviour is very different to that for the case of a rigid outer boundary for which the corresponding form for $\bar{\mathring \omega}\bigl(r^\star/H,E\Omega t^\star\bigr)$ decays (see the $n=1$ term of (\ref{QG-me-GH}{$a$})) rather than grow exponentially. The striking contrast made in our discussion \S\ref{Discussion} between (\ref{ms-as-sol}) and (\ref{ms-as-sol-GH}), which incorporate these results, highlights the central theme of our paper.

\subsection{Outline\label{Outline}}

Our notation is complicated and best outlined in general terms at the outset. Our basic variables, like the angular velocity are identified by the breve accent $\breve{\,\,}\,$. Only variables with a superscript star $^\star$ are dimensional; otherwise they are dimensionless. Azimuthal QG-flows independent of $z^*$ are identified by an overline $^{\bar{\,\,}}\,$. All these notations were introduced in \S\ref{GH63-problem}. There the mathring accent $\mathring{\,\,}$ was also introduced in the restricted sense of mainstream QG-flows, being the amplitude that remains after the spin-down decay factor $\exp\bigl(-\,\sigma E^{1/2}\Omega t^\star\bigr)$ is removed (see (\ref{Ho-Gr})). Below, we will however use it in the general sense (see (\ref{sol-tau}$a$)), which includes the mainstream QG-part together with the Ekman layer contribution denoted by the tilde $\tilde{\,\,}\,$ (see (\ref{MS-EL})). The final eigensolution that emerges as $t^\star\to \infty$, like the QG-form (\ref{ms-lt}) with the decay factor $\exp\bigl(-\,qE\Omega t^\star\bigr)$, is otherwise without accents (see (\ref{sol-tau-final})).

We formulate the transient flow problem in its entirety in \S\ref{mathematical-problem}. We outline the basic properties of the Ekman layer together with its consequences for the QG-flow in  \S\ref{Ekman-Layer}, particularly the QG-momentum equation (\ref{QG-eq-b}) governing ${\bar{\mathring \omega}}$ and its relation (\ref{MS-flux}$a$) to the $z^\star$-average $\langle{\mathring \omega}\rangle$. They differ by a small amount due to the angular momentum deficit in the Ekman layer. Though the difference is small, we find that our comparisons with the numerics and asymptotics is improved by use of (\ref{MS-flux}$a$). We consider the eigenvalue problem for the final decay of the QG-flow (\ref{ms-lt}$a$) in \S\ref{QG-final-decay} from an asymptotic point of view on the basis that the eigenvalue $k$ in (\ref{ms-lt}$b$), related to $\alpha$ in (\ref{ms-bc}), is known. Then in \S\ref{DNS} we report results from direct numerical simulations, which confirm that the solution approaches the proposed asymptotic form. In \S\ref{ENS}, we identify the shape and decay rate but not the amplitude of the eigensolution again using  numerical methods.

The transient evolution from the classical spin-down mainstream flow (\ref{simple-spin-down}) to the final decay mode (\ref{ms-lt}) (or more specifically (\ref{ms-as-sol})) is of particular interest. So in \S\ref{QG-trans} we solve the angular momentum equation (\ref{QG-eq-b}) for $\bar{\mathring \omega}$ subject to the initial condition (\ref{ms-ic}) on the basis that $\alpha$ in the boundary condition (\ref{ms-bc}) is known. The numerical solution is described in \S\ref{QG-num-res}, while a power series solution valid for short times is described in \S\ref{QG-series} which notably is valid on the spin-down time $E^{-1/2}/\Omega$ when the shear-layer width is $E^{1/4}H$. On that time it provides our analogue of the rigid boundary solution \citep[][eq.~(6.3)]{GH63}. Indeed our power series is informative on much of the lateral diffusion time scale $\ell^2 E^{-1}/\Omega$ and even provides a good approximation to $\bar{\mathring \omega}$ at the outer boundary until the final asymptotic behaviour is clearly evident.

For a complete understanding of the transient evolution together with the final decay mode, we need a theory for $\alpha$. To that end, in \S\ref{side-wall} we formulate the boundary layer equations for the $E^{1/3}$-layer correct not simply to leading order but correct to $O(E^{1/6})$. We study the leading order problem asymptotically in \S\ref{0-order} in the spirit of \cite{S57}. As the solution is singular near the bottom outside corner at $(r^\star,z^\star)=(L,0)$, there it is best expressed in a similarity form which we provide and expand upon in \S\ref{0-ps}. Unfortunately this leading order solution contains no QG-part. To extract the QG-part, we continue in \S\ref{side-wall-QG} to $O(E^{1/6})$ at which the singularity forces a $\ln E$ type dependence in the asymptotic value (\ref{sl-ln-E}) of $\alpha$. As the numerical value of $E$ needed for the applicability of the asymptotic theory is minute, we cannot compare directly asymptotic and  numerical results. To overcome this obstacle  we study in \S\ref{side-wall-hybrid} the numerical solution of the side-wall boundary layer equations (\ref{sl-eq}), containing the parameter $\epsilon\equiv E^{1/6}$ explicitly at fixed $E$. Their numerical solution determines $\alpha$, which we find agrees with that predicted by the full numerical eigensolution. In summary, we are unable to solve numerically the complete governing equations (\ref{mom-vort}) at values of $E$ small enough to reach the true asymptotic limiting behaviour described in  \S\ref{side-wall-asym} which is applicable in the limit $E\to 0$. However, our numerical solutions of the asymptotically derived boundary layer equations (\ref{sl-eq}) in  \S\ref{side-wall-hybrid} together with the transient results of \S\ref{QG-trans} lead to a comprehensive and consistent picture of the asymptotic behaviour of the solution at small but finite $E$.

\bigskip

\section{The mathematical problem\label{mathematical-problem}}

Whereas our primary concern is with the evolution of $\bar{{\breve \omega}^\star}$ on the long lateral diffusion time-scale $\ell^2E^{-1} /\Omega$ together with the final eigenmode (\ref{ms-lt}) state, we will non-dimensionalise our governing equations on the short inertial wave time-scale $1/\Omega$ as that is the time-scale on which inertial waves are manifest in the transient solution. Our numerical solution needs to take that evolution into account. Accordingly on writing
\bme
\label{dim-var}
\be
\te
(r^\star,\,\theta^\star,\,z^\star)\,=\,(Hr,\,\theta,\,Hz)\,, \qquad\quad {\breve \uv}^\star\,=\,\varepsilon H \Omega {\breve \uv} \,, \qquad\quad
t^\star\,=\,t/\Omega\,,
\ee
\eme
the equations of motion are
\bme
\label{gov-eq}
\be
\partial {\breve \uv}/\partial t\,+\,2\vp{\hat \zv}{\breve \uv}\,=\,\bfnabla {\breve p}\,+\,E\nabla^2{\breve \uv}\,,\qquad\qquad
\sp{\bfnabla}{{\breve \uv}}\,=\,0\,,
\ee
where ${\hat \zv}$ is the unit vector in the $z$-direction ($\Omegav\equiv\Omega{\hat \zv}$) and ${\breve p}$ is a suitably non-dimensionalised measure of pressure. The corresponding vorticity equation is
\be
\se
\partial (\vp{\bfnabla}{\breve \uv})/\partial t\,-\,2\sp{\hat \zv}{\bfnabla}{\breve \uv}\,=\,E\nabla^2(\vp{\bfnabla}{\breve \uv})\,.
\ee
\eme
We write
\bme
\label{psi-omega}
\se
\begin{align}
{\breve \uv}\,=\,\bigl({\breve u}\,,\,{\breve v}\,,\,{\breve w}\bigr)
     =&\,\biggl(-\,\dfrac{E^{1/2}}{r}\pd{\breve \psi}z\,,\,r{\breve \omega}\,,\,\dfrac{E^{1/2}}{r}\pd{{\breve \psi}}r\biggr),\\[0.3em]
\vp{\bm \nabla}{\breve \uv}\,=&\,\biggl(-\,{r}\pd{{\breve \omega}}z\,,E^{1/2}{\breve \gamma}\,,\,\dfrac{1}{r}\pd{\,}r(r^2{\breve \omega})\biggr) ,
\end{align}
where
\be
\de
{\breve \gamma}\,=\,-\,\Bigl(\nabla^2\,-\,\dfrac{1}{r^2}\Bigr)\Bigl(\dfrac{\breve \psi}{r}\Bigr)\,=\,-\,\dfrac1r\DC{\breve \psi}\,,\qquad\qquad
\DC\,=\,r\pd{\,}r\Bigl(\dfrac{1}{r}\pd{\,}r\Bigr)+\,\pd{^2\,}{z^2}\,.
\ee
\eme
The scaling of the streamfunction ${\breve \psi}$ by the factor of $E^{1/2}$ anticipates our primary interest in QG-flows driven by Ekman suction. The azimuthal components of the momentum and vorticity equations (\ref{gov-eq}) give
\bse
\label{mom-vort}
\begin{align}
r^2\,\pd{\breve \omega}{t}\,-\,2\,E^{1/2}\pd{{\breve \psi}}z\,&=\,E\,\DC\bigl(r^2{\breve \omega}\bigr)\,,\\[0.3em]
\pd{\,}{t}\bigl(\DC{\breve \psi}\bigr)+\,\dfrac{2r^2}{E^{1/2}}\,\pd{{\breve \omega}}z\,&=\,E\DC^2{\breve \psi}\,.
\end{align}
\ese
The initial conditions are
\be
\label{ic}
{\breve \psi}\,=\,0\,,\qquad\qquad {\breve \omega}\,=\,1 \qquad\qquad\mbox{everywhere at} \qquad t=0\,.
\ee
For $t>0$ the boundary conditions are
\bse
\label{bc}
\begin{align}
{\breve \psi}\,=\,\pd{{\breve \omega}}{r}\,=\,\pd{{\breve w}}{r}\,&=\,0& \mbox{at} &&r\,=&\,0\,\,\,\mbox{ and }\,\,\ell & (0&<z<1)\,,\\
{\breve \psi}\,=\,\pd{{\breve \psi}}{z}\,=\,{\breve \omega}\,&=\,0&  \mbox{at}&& z\,=&\,0 & (0&<r<\ell)\,,\\
\qquad{\breve \psi}\,=\,\pd{^2{\breve \psi}}{z^2}\,=\,\pd{{\breve \omega}}{z}\,&\,=\,\,0&  \mbox{at} &&z\,=&\,1 &(0&<r<\ell)\,.\qquad
\end{align}
\ese
 
In order to discuss the evolution on the longer lateral diffusion time $\ell^2E^{-1}/\Omega$, we make the further change of variables
\bme
\label{sol-tau}
\be
\bigl[{\breve \uv}\,,\,{\breve \psi}\,,\,{\breve \omega}\bigr]\,=\,\kappa\bigl[{\mathring \uv}\,,\,{\mathring \psi}\,,\,{\mathring \omega}\bigr](r,z,\tau)\,\exp(-\sigma E^{1/2}t) \,, \qquad\quad  t\,=\,\ell^2E^{-1}\tau\,,
\ee
\eme
where we have incorporated the amplitude change $\kappa$ and decay rate $\sigma E^{1/2}$ (see (\ref{simple-spin-down}$b$,$c$) respectively), predicted by the spin-down solution described in appendix~\ref{infinite-layer} to occur on the shorter time $E^{-1/2}/\Omega$ (see also (\ref{omegabar})). The new variables ${\mathring \uv}$, ${\mathring \psi}$ and ${\mathring \omega}$ satisfy (\ref{gov-eq})--(\ref{bc}) as before but with
\be
\label{two-timing}
\pd{\,}{t} \,\mapsto \,  \ell^{-2}E\pd{\,}{\tau}\,-\,\sigma E^{1/2}\,.
\ee
So, for example, (\ref{mom-vort}$a$,$b$) become
\bse
\label{mom-vort-l-t}
\begin{align}
\ell^{-2}E^{1/2}r^2\,\pd{\mathring \omega}{\tau}\,-\,\sigma r^2{\mathring \omega}\,-\,2\,\pd{{\mathring \psi}}z\,&=\,E^{1/2}\,\DC\bigl(r^2{\mathring \omega}\bigr)\,,\\[0.3em]
\ell^{-2}E^{1/2}\pd{\,}{\tau}\bigl(\DC{\mathring \psi}\bigr) \,-\,\sigma\DC{\mathring \psi}\,+\,\dfrac{2r^2}{E}\,\pd{{\mathring \omega}}z\,&=\,E^{1/2}\DC^2{\mathring \psi}\,.
\end{align}
\ese

We decompose the velocity ${\mathring \uv}$ into its mainstream $\bigl(\,\bar{\mathring \uv}\,\bigr)$ and Ekman layer $\bigl(\tilde{\mathring \uv}\bigr)$ parts:  
\be
\label{MS-EL}
{\mathring \uv}\,=\,\bar{\mathring \uv}\,+\,\tilde{\mathring \uv}\,,
\ee
with similar decompositions for the other variables. The QG-nature of the mainstream flow is dictated by (\ref{mom-vort-l-t}$b$), which with (\ref{mom-vort-l-t}$a$) determines
\bme
\label{MS-sol}
\be
{\mathring \omega}\,=\,{\bar{\mathring \omega}}(r,\tau)\, +\,\OR(E)\,,\qquad\qquad
\left\{\begin{array}{l}
{\mathring \psi}\,=\,{\bar{\mathring \psi}}(r,\tau)\,(z-1)\,+\,\OR(E)\,,\\[0.3em]
{\mathring u}\,=\,{\bar{\mathring u}}(r,\tau)\,+\,\OR(E^{3/2})\,,\end{array}\right.
\ee
where, from (\ref{psi-omega}$a$), (\ref{MS-sol}$b$) and (\ref{mom-vort-l-t}$a$) again,
\be
\se
-\,{\bar{\mathring \psi}}\,=\,E^{-1/2}r{\bar {\mathring u}}\,=\,\tfrac12 \biggl[\sigma r^2{\bar{\mathring \omega}}\,-\,E^{1/2}\biggl(\dfrac{r^2}{\ell^2}\,\pd{\bar{\mathring \omega}}\tau\,-\,\DC (r^2{\bar{\mathring \omega}})\biggr)\biggr]
\ee
\eme
and ${\bar{\mathring \omega}}(r,\tau)$ is a function of $r$ and $\tau$ yet to be determined. 

A fortunate feature of our two time-scale ansatz is that (\ref{MS-sol}$c$) and (\ref{mom-vort-l-t}$b$) are solved, correct to $O(E^{1/2})$ as advocated in (\ref{MS-sol}$a$,$b$), by
\be
\label{QG-eq-a}
-\,{\bar{\mathring \psi}}\,=\,E^{-1/2}r{\bar {\mathring u}}\,=\,\tfrac12 \sigma r^2{\bar{\mathring \omega}}\,,
\ee
\be
\label{QG-eq-b}
\pd{\bar{\mathring \omega}}\tau\,=\,\dfrac{\ell^2}{r^2}\DC (r^2{\bar{\mathring \omega}})\,.
\ee
Equation (\ref{QG-eq-b}) highlights the absence of any $E^{1/4}$-Stewartson layer, usually triggered by the Ekman suction term $-{\bar{\mathring \psi}}$ on the left-hand side of (\ref{MS-sol}$c$). For our spin-down flow, that effect is exactly balanced by the inertial decay $\tfrac12 \sigma r^2{\bar{\mathring \omega}}$ on the right-hand side (see (\ref{QG-eq-a})). 

\subsection{Ekman Layer\label{Ekman-Layer}}

In the Ekman layer near $z=0$ we set
\bme
\label{EL-variables}
\be
{\mathring v}\,=\,r{\bar{\mathring \omega}}(r,\tau)\,+\,{\tilde{\mathring v}}\,,\qquad\qquad
\left\{\begin{array}{l}
{\mathring \psi}\,=\,{\bar {\mathring \psi}}(r,\tau)\,(z-1)\,+\,{\tilde {\mathring \psi}}\,,\\[0.3em]
{\mathring u}\,=\,{\bar{\mathring u}}(r,\tau)\,+\,{\tilde{\mathring u}}\,.\end{array}\right.
\ee
\eme
Correct to leading order, the governing equation (\ref{gov-eq}), as modified in (\ref{mom-vort-l-t}), determines the Ekman layer equations
\bme
\label{EL-eq}
\be
\left.\begin{array}{l}
-\,{E^{1/2}\sigma}{\tilde {\mathring v}}\,+\,2{\tilde {\mathring u}}\,=\,E\pd{^2{\tilde {\mathring v}}}{z^2}\,,\\[0.8em]
-\,{E^{1/2}\sigma}{\tilde {\mathring u}}\,-\,2{\tilde {\mathring v}}\,=\,E\pd{^2{\tilde {\mathring u}}}{z^2}\,,\end{array}\right\}
\qquad\qquad
\left\{\begin{array}{l}
r{\tilde {\mathring u}}\,=\,-\,E^{1/2}\pd{\tilde {\mathring \psi}}z\,,\\[0.8em]
E^{1/2}{\tilde {\mathring \psi}}\,+\,r{\bar {\mathring u}}\,=\,-r{\ds\int_0^z}{\tilde {\mathring u}}\,\dR z\,,
\end{array}\right.
\ee
\eme
where note has been taken of (\ref{MS-sol}{$b$}), as well as $E^{1/2}{\tilde{\mathring \psi}}=E^{1/2}{\bar{\mathring \psi}}=-r{\bar{\mathring u}}$ (see (\ref{QG-eq-a})) at $z=0$. Equations~(\ref{EL-eq}$a$,$b$) must be solved subject to ${\tilde {\mathring v}}=-r{\bar{\mathring \omega}}$, ${\tilde {\mathring u}}=-{\bar{\mathring u}}$ at $z=0$ with ${\tilde {\mathring v}}$, ${\tilde {\mathring \psi}}$, ${\tilde {\mathring u}}$ all tending to zero as $z/E^{1/2}\uparrow\infty$. We rewrite (\ref{EL-eq}a) compactly as
\be
\label{EL-eq-Z}
E\pd{^2Z}{z^2}\,-\bigl(2\iR -\,E^{1/2}\sigma\bigr)Z\,=\,0\,,\qquad\qquad  Z\,=\,{\tilde {\mathring v}}\,-\,\iR\,{\tilde {\mathring u}}\,,
\ee
with solution
\be
\label{EL-sol-Z}
Z\,=\,-\bigl(r{\bar{\mathring \omega}}-  \iR {\bar {\mathring u}}\bigr)\exp\Bigl[-E^{-1/2}\bigl(1+\tfrac12\iR E^{1/2}\sigma\bigr)^{1/2}(1+\iR)z\Bigr].
\ee
From (\ref{EL-eq}b) evaluated as $E^{-1/2}z\uparrow\infty$ we obtain
\be
\label{EL-sol-MS}
r{\bar {\mathring u}}\,=\,-\,r \Re\biggl\{\int_0^\infty \iR Z\,\dR z\biggr\}\,=\,\tfrac12 E^{1/2}r\,\Re\Biggl\{\dfrac{\bigl(r{\bar{\mathring \omega}}-  \iR {\bar {\mathring u}}\bigr)(1+\iR)}{\bigl(1+\tfrac12\iR E^{1/2}\sigma\bigr)^{1/2}}\Biggr\}.
\ee
Using (\ref{QG-eq-a}), namely ${\bar {\mathring u}}=\tfrac12\sigma E^{1/2}r{\bar{\mathring \omega}}$, we recover the formula (\ref{p-sigma}) for $\sigma$.

Also of interest to us is the azimuthal Ekman layer flux 
\be
\label{EL-flux}
\int_0^\infty r{\tilde{\mathring \omega}}\,\dR z\,=\,\Re\biggl\{\int_0^\infty Z\,\dR z\biggr\}\,=\,-\,\tfrac12 E^{1/2}r{\bar{\mathring \omega}}\,+\,O(E)\,.
\ee
Using (\ref{EL-variables}$a$) it means that the $z$-average of ${\mathring \omega}(r,z,\tau)$ is
\bme
\label{MS-flux}
\be
\langle{\mathring \omega}\rangle\,=\,\bigl(\mu\,+\,O(E)\bigr){\bar{\mathring \omega}}(r,\tau)\,,
\qquad\qquad \mu\,=\,1-\tfrac12 E^{1/2}\,+\,O(E)
\ee
(see (\ref{kappa-mu}$b$)), where
\be
\se
\langle\,\cdots\,\rangle\,\equiv\,\int_0^1 \cdots\,\dR z
\ee
\eme
and $\mu$ has the specific definition (\ref{inf-layer-sol-av}$b$). So, in the case of ${\bar{\mathring \omega}}=\,$const.~considered in appendix~\ref{infinite-layer}, the relation $\langle{\mathring \omega}\rangle=\mu{\bar{\mathring \omega}}$ holds exactly without the $O(E)$ correction mentioned in (\ref{MS-flux}$a$). We stress this subtle difference as our Ekman layer calculation does not consider the $O(E)$ Ekman layer corrections due to the effect of the term $E(\nabla^2-\partial^2/\partial z^2)\tilde{\mathring \uv}$ ignored in (\ref{EL-eq}$a$,$b$). Whereas $\bar{\mathring \omega}$ is the natural quantity to consider from the point of view of asymptotics, only $\langle{\mathring \omega}\rangle$ can be measured unambiguously from our direct numerical simulation (DNS) of the complete governing equations (\ref{mom-vort}) at finite $E$. For the largest value $E=3\times 10^{-4}$ used, the small $O\bigl(E^{1/2}\bigr)$ difference is not insignificant and taking the correction into account improves the accuracy of our comparison of the DNS results with the asymptotics.

\subsection{The final decay of the QG-flow: $\tau \to \infty$\label{QG-final-decay}}

As pointed out earlier, the $r=\ell$ mainstream boundary condition (\ref{ms-bc}) only emerges after proper consideration of the $E^{1/3}$--side-wall boundary layer, undertaken in \S\ref{side-wall} below. Here we simply note that the final decay mode is described by (\ref{ms-lt}$a$) which relative to the spin-down decay grows and takes the form (\ref{ms-lt-1}):
\be
\label{QG-decay}
\Bigl[\,\bar{\mathring \psi}\,,\,\bar{\mathring \omega}\,\Bigr](r,\tau)\,\approx \bigl[\,\bar{\psi}\,,\,\bar{\omega}\,\bigr](r)\,\exp\bigl(k^2\tau\bigr)\qquad\qquad \mbox{as} \qquad\qquad \tau\to \infty\,.
\ee
It satisfies (\ref{QG-eq-b}) when
\be
\label{Bessel-eq}
r\od{\,}r\biggl(r\od{\,}r(r{\bar\omega})\biggr)-\biggl(\Bigl(\dfrac{k r}{\ell}\Bigr)^2+1\biggr)r{\bar\omega}\,=\,0\,.
\ee
In terms of the Modified Bessel function $\IR_1$, the solution regular at $r=0$ is
\be
\label{Bessel-sol}
{\bar\omega}(r)/A_0\,=\,{\bar{\mathring \omega}}_0(r)\,\equiv\,2(k r/\ell)^{-1}\,\IR_1(k r/\ell)\,,
\ee
where the normalisation constant $A_0$ is chosen (see (\ref{Bessel-A0}) and (\ref{left-right-end-values}) below) such that ${\bar{\mathring \omega}}_0(0)=1$. From (\ref{Bessel-sol}) we deduce that
\bme
\label{kalpha}
\begin{align}
\ell\od{{\bar{\mathring \omega}}_0}r(\ell)\,=\,2k\biggl(\dfrac{\IR_1(\rho)}{\rho}\biggr)^{\!\prime}\,\biggr|_{\rho=k}\,=\,2\IR_2(k)\,\equiv\,\,&\,{\mathring \alpha}\,{\bar{\mathring \omega}}_0(\ell)\,,\qquad
& {\mathring \alpha}\,=\,&\,k\,\dfrac{\IR_2(k)}{\IR_1(k)}\,,\\
\mbox{equivalently} \qquad\qquad
\od{\bar\omega}r(\ell)\,=\,&\,\alpha\,{\bar\omega}(\ell)\,,\qquad  &
\alpha\,=\,&\,{\mathring \alpha}/\ell\,,
\end{align}
\eme
where the prime denotes differentiation. Consideration of the transient QG-evolution in \S\ref{QG-trans} below shows that
\be
\label{Bessel-A0}
A_0\,=\,\tfrac12 k^2 \IR_2(k)\bigg/\!\! \int_0^k\! \rho \bigl[\IR_1(\rho)\bigr]^2\,\dR\rho
\qquad\qquad
\biggl(k^2 \IR_2(k)\,=\int_0^k \!\rho^2 \IR_1(\rho)\,\dR\rho\biggr)
\ee
(see (\ref{QG-asym})). Integration of the identity
\bse
\label{Bessel-identity}
\be
\se
2\rho\bigl[\IR_1(\rho)\bigr]^2\,=\,\Bigl\{-\,\bigl[\rho \IR_1^{\,\prime}(\rho) \bigr]^2 \,+\,\bigl(1+\rho^2\bigr)\bigl[\IR_1(\rho) \bigr]^2\Bigr\}^{\prime}
\ee
and noting again that $\rho \IR_1^{\,\prime}(\rho)=\rho\IR_2(\rho)+\IR_1(\rho)$ determines
\be
2\int_0^k\! \rho \bigl[\IR_1(\rho)\bigr]^2\,\dR\rho\,=\,\bigl(k^2\,-\,2{\mathring \alpha}\,-\,{\mathring \alpha}^2\bigr)\bigl[\IR_1(k)\bigr]^2
\ee
\ese
on use of (\ref{kalpha}$b$). Substitution of (\ref{Bessel-identity}$b$) into (\ref{Bessel-A0}) evaluates $A_0$ and whence from (\ref{Bessel-sol}) we determine the boundary values
\be
\label{left-right-end-values}
\bar{\mathring \omega}(r,\tau)\exp\bigl(-k^2\tau\bigr)
\,\to\,\left\{\begin{array}{ll}
A_0 \,\equiv\,\dfrac{k{\mathring \alpha}}{k^2\,-\,2{\mathring \alpha}\,-\,{\mathring \alpha}^2}\,\dfrac1{\IR_1(k)}\qquad&\mbox{at} \qquad r=0\,,\\[1.0em]
A_\ell\equiv \,\dfrac{2{\mathring \alpha}}{k^2\,-\,2{\mathring \alpha}\,-\,{\mathring \alpha}^2}\qquad&\mbox{at}\qquad r=\ell\,.
\end{array}\right.
\ee
Essentially in (\ref{Bessel-sol}) we have identified the first term of a Dini series modal expansion (see \cite{EMOT53} \S7.10.4, especially eq.~(49), and \cite{W66} chapter XVIII).

Finally to complete the solution (\ref{QG-decay}), we note that (\ref{QG-eq-a}) determines
\be
\label{MS-sol-new}
-\,{\bar\psi}\,=\,E^{-1/2}r{\bar u}\,=\,\tfrac12 \sigma r^2{\bar\omega}\,=\,\tfrac12\bigl(1+\tfrac34 E^{1/2}+O(E)\bigr) r^2{\bar\omega}\,.
\ee

\subsection{Direct numerical simulation (DNS) for the case $\ell=1$, $E=10^{-4}$ \label{DNS}}

\begin{figure}
\centerline{}
\vskip 3mm
\centerline{
\includegraphics*[width=1.0 \textwidth]{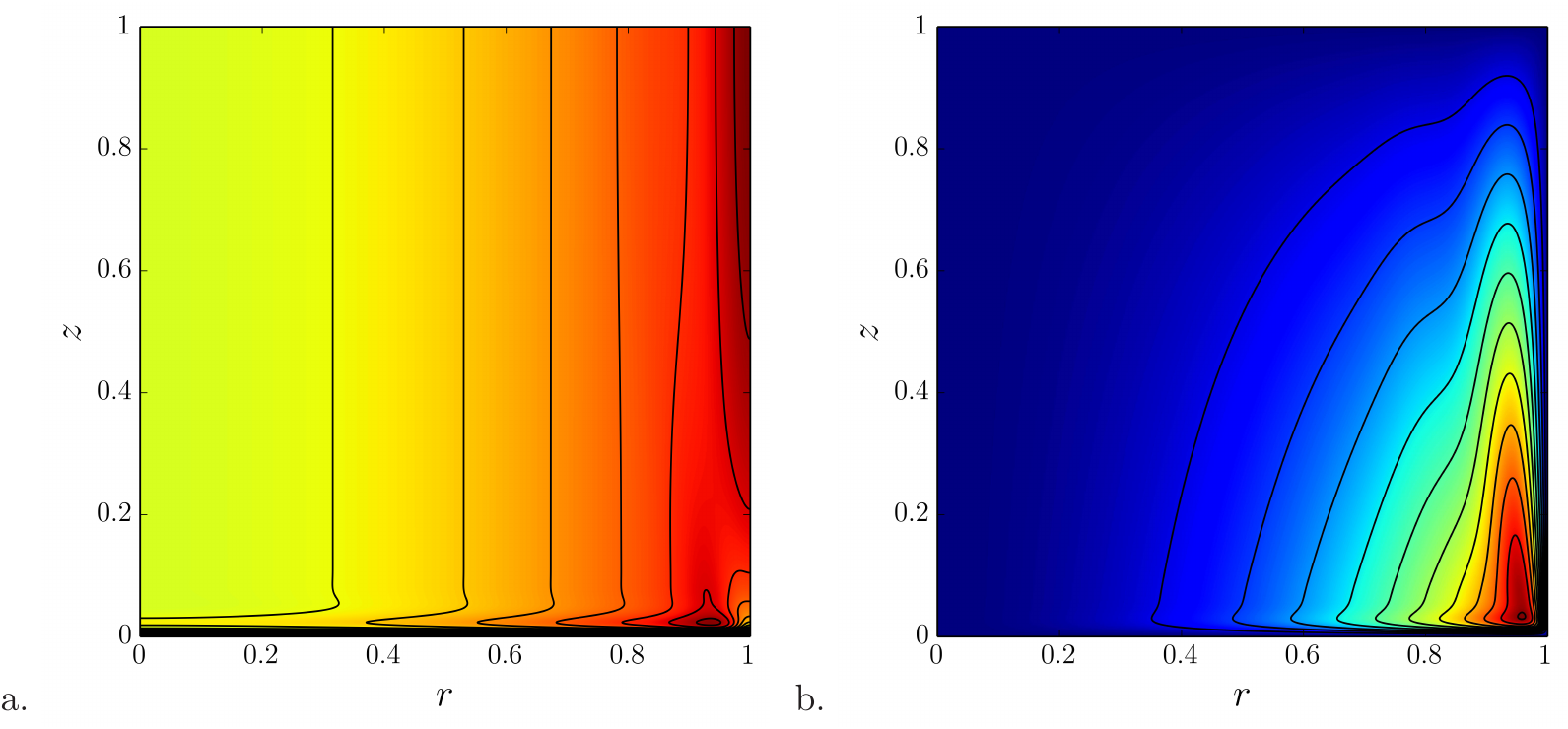}
}
\caption{(Colour online) The contours in the $r$--$z$ plane of ($a$) the angular velocity $\omega_\tDNS$ and ($b$) the streamfunction $\psi_\tDNS$, obtained from the DNS of the initial value problem (\ref{mom-vort})--(\ref{bc}) for the case $\ell=1$, $E=10^{-4}$ in the large $\tau$ limit (see (\ref{sol-tau-final})).}
\label{EX}
\end{figure}

We performed DNS of the governing equations (\ref{mom-vort}) subject to the initial conditions (\ref{ic}) and boundary conditions (\ref{bc}). We will distinguish such solutions by the subscript $\tDNS$, i.e., our DNS-solution is $\bigl[{\breve \uv}\,,\,{\breve \psi}\,,\,{\breve \omega}\bigr]=\bigl[{\breve \uv}_\tDNS\,,\,{\breve \psi}_\tDNS\,,\,{\breve \omega}_\tDNS\bigr](r,z,t)$. We solved (\ref{mom-vort}) using second-order finite differences in space, and an implicit second-order backward differentiation (BDF2) in time. We used a stretched grid, staggered in the $z$-direction. Each simulation was initialised with a uniform distribution of ${\breve{\omega}}_\tDNS$. The spatial resolution depends on the value of $E$, and was varied up to $2000 \times 2000$ to ensure convergence.

As $\tau\to\infty$ the final eigensolution takes the form 
\be
\label{sol-tau-final}
\kappa^{-1}\bigl[{\breve \uv}_\tDNS\,,\,{\breve \psi}_\tDNS\,,\,{\breve \omega}_\tDNS\bigr]\,=\,\bigl[{\uv}_\tDNS\,,\,{\psi}_\tDNS\,,\,{\omega}_\tDNS\bigr](r,z)\,\exp(-q_\tDNS\ell^2\tau)\,,
\ee
similar to the analytic prediction (\ref{ms-lt}$a$,$b$). The contours of constant $\psi_\tDNS$ and $\omega_\tDNS$ for the case $\ell=1$, $E=10^{-4}$, for which $q_\tDNS\approx 96.9648$, are illustrated in figure~\ref{EX}. A thin Ekman layer is visible near $z=0$, linked to the Ekman layer contributions $\tilde \psi$ and $\tilde \omega$, identified by ${\tilde {\mathring \psi}}$ and ${\tilde {\mathring \omega}}$ as $\tau\to \infty$. The QG-nature of the mainstream is demonstrated by the $\omega_\tDNS$--contours, which are almost parallel to the $z$-axis in figure~\ref{EX}($a$). The mainstream outflow caused by Ekman blowing is revealed by the tilted $\psi_\tDNS$--contours in figure~\ref{EX}($b$), whose nature is consistent with the mainstream asymptotic result  
\be
\label{psi-QG}
\psi\,=\,(z-1){\bar \psi}(r)\,\approx\,-\,\tfrac12\sigma (z-1)r^2{\bar \omega}(r) 
\ee
determined by (\ref{MS-sol}$b$) and (\ref{MS-sol-new}). This outflow is returned inside the $E^{1/3}$--side-wall layer (see \S\ref{side-wall}) , evident near $r=\ell(\,=1)$, to a sink, namely the $E^{1/2}\times E^{1/2}$ corner region in the neighbourhood of $(r,z)=(\ell, 0)$. In turn that influx is ejected within the $E^{1/2}$ Ekman layer at the base of the $E^{1/3}$-layer to provide the Ekman layer flux needed to adjust the QG-flow.

\subsection{Numerical eigensolution (ENS) for the case $\ell=1$ for various $E$ \label{ENS}}

\begin{figure}
\centerline{}
\vskip 3mm
\centerline{
\includegraphics*[width=1 \textwidth]{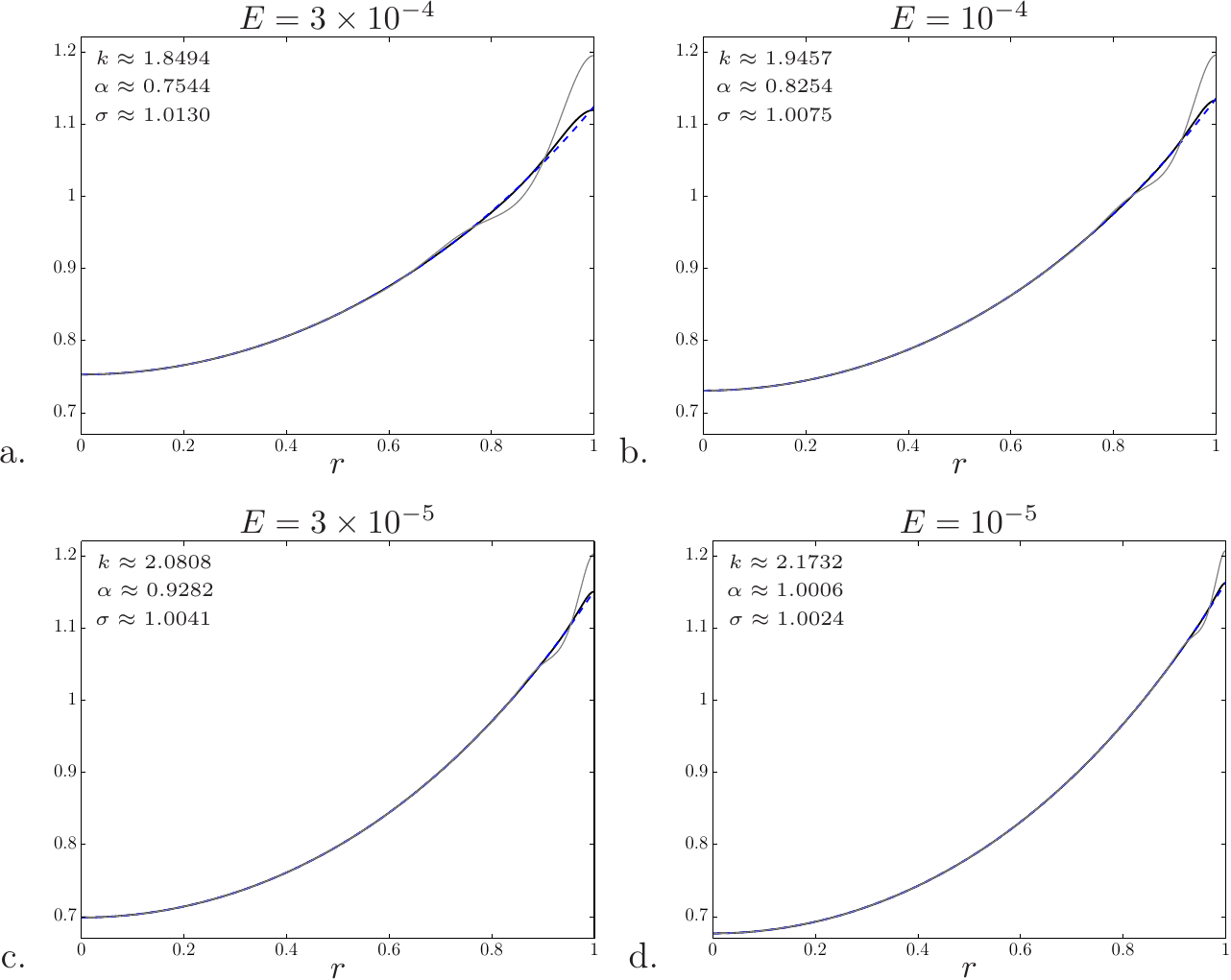}
}
\caption{Comparison of ENS results ${\bar\omega}_{\tENS}(r)$ (see (\ref{omega-ENS}): continuous dark line) and $\omega_{\tENS}(r,1)$ (continuous light line), normalised by ${\bar\omega}_{\tENS}(0)={\bar\omega}(0)$, with the entire analytic mainstream solution ${\bar \omega}(r)$ (dashed line) on $0\le r\le \ell\,(=1)$. Plots are for the cases: ($a$) $E=3\times 10^{-4}$; ($b$) $E=10^{-4}$; ($c$) $E=3\times 10^{-5}$; ($d$) $E=10^{-5}$. The values of $k$, $\alpha$ and $\sigma$ for each case are given on the respective figures.}
\label{figk}
\end{figure}
As the DNS eigensolution (\ref{sol-tau-final}) requires very considerable computational time, we found it more efficient to extract the functional form without detailed consideration of the transient evolution. By that expedient we were able to solve the eigenvalue problem for $\uv$ (but only up to an arbitrary constant) and $q$ (i.e., $\uv_\tDNS$ and $q_\tDNS$ in (\ref{sol-tau-final})) for the case $\ell=1$ for various values of $E$. We refer to it as the eigen (numerical) solution (ENS), for which  $[\uv\,,\,\psi\,,\,\omega]=\bigl[\uv_\tENS\,,\,\psi_\tENS\,,\,\omega_\tENS\bigr](r,z)$ and $q= q_\tENS$. To place the results in context we note that our asymptotic theory predicts
\be
\label{MS-flux-extra}
\bigl\langle{\breve \omega}\bigr\rangle(r,t)\,\approx\,\mu{\bar{\breve \omega}}(r,t)\,\approx\,(\mu\kappa)\,{\bar\omega}(r)\,\exp(-q\ell^2\tau)\qquad\qquad\mbox{as}\qquad \tau\to\infty
\ee
(see (\ref{sol-tau}), (\ref{MS-flux}), (\ref{QG-decay}) and (\ref{sol-tau-final})), where ${\bar\omega}(r)$ is defined by (\ref{Bessel-sol}). Since the notion of a mainstream QG-flow and an Ekman layer is an asymptotic ($E\ll 1$) concept, ${\bar\omega}_{\tENS}(r)$ evaluated at finite $E$ is not clearly defined. To overcome this obstacle we simply set
\be
\label{omega-ENS}
{\bar\omega}_{\tENS}(r)\,\equiv \,\mu^{-1}\langle \omega_{\tENS}\rangle(r)
\ee
so that comparisons can be made with asymptotic theory for which $q$ in (\ref{MS-flux-extra}) is unfortunately as yet unknown. However, on assuming that $q=q_\tENS$ we may infer that the value of $k$ characterising our analytic solution ${\bar\omega}(r)$ (see (\ref{Bessel-sol})) is given by $k^2=\ell^2\bigl(\sigma E^{-1/2}-q_\tENS\bigr)$ (see (\ref{ms-lt}$b$)). As the amplitude of the eigenfunction $\omega_{\tENS}(r,z)$ is arbitrary, we simply normalise it by ${\bar\omega}_{\tENS}(0)={\bar\omega}(0)$. If the ENS is truly QG we expect the upper boundary value $\omega_{\tENS}(r,1)$ to equal  ${\bar\omega}(r)$. For that reason we compare the plots of ${\bar\omega}(r)$, ${\bar\omega}_{\tENS}(r)$ and $\omega_{\tENS}(r,1)$ in figures~\ref{figk}($a$--$d$) for various values of $E$. The plots confirm our mainstream expectations: ${\bar\omega}(r)\approx {\bar\omega}_{\tENS}(r)\approx \omega_{\tENS}(r,1)$. Their values only differ significantly within the $E^{1/3}$--side-wall ($r=\ell$) layer. The strength of that layer is indicated by the amplitude of the oscillations $\omega_{\tENS}(r,1)-{\bar\omega}(r)=O\bigl(E^{1/6}\bigr)$. The magnitude of the oscillations ${\bar\omega}_{\tENS}(r)-{\bar\omega}(r)=O\bigl(E^{1/3}\bigr)$ are markedly smaller. The relative sizes are clearly visible on the plots, while the theory behind the orders of magnitude alluded to will be explained in \S\ref{side-wall}. 

We emphasise that so far the decay rate $q$ has been predicted by the ENS. To obtain a closed form analytic solution we need to determine the value of $\alpha$ that characterises the mainstream boundary condition $\dR{\bar \omega}/\dR r(\ell)=\alpha {\bar \omega}(\ell)$ (see (\ref{kalpha}$c$)) by consideration of the $E^{1/3}$--side-wall layer. With $\alpha$ known, the $\tau$-growth rate $k^2$ (see (\ref{QG-decay})) is given by the solution $k=k({\mathring \alpha})$ of $(k/\ell)\IR_2(k)/\IR_1(k)={\mathring \alpha}=\ell\alpha$ (see (\ref{kalpha}$b$,$d$)). That strategy, attempted in  \S\ref{side-wall-asym}, is hindered by the asymptotic orderings ${\omega}(r,z)-{\bar\omega}(r)=O\bigl(E^{1/6}\bigr)$, $\mu^{-1}\langle \omega\rangle(r)-{\bar\omega}(r)=O\bigl(E^{1/3}\bigr)$ hinted at by figures~\ref{figk}($a$--$d$). In short, though we are able to solve the vital $O\bigl(E^{1/3}\bigr)$ problem for the relatively small shear-layer correction $\mu^{-1}\langle \omega\rangle-{\bar\omega}$, the tiny size of $E$ needed for its validity is unreachable by our ENS. To bypass this difficulty, in \S\ref{side-wall-hybrid} we consider the shear-layer problem numerically retaining the relevant terms involving $E$ needed to encompass both the $O\bigl(E^{1/6}\bigr)$ and $O\bigl(E^{1/3}\bigr)$ problems simultaneously. From this hybrid asymptotic-numerical method  we determine $\alpha$ and in turn ${\mathring \alpha}=\ell\alpha$  and  $k=k({\mathring \alpha})$ for various values of $E$. The ENS value $q_\tENS$ and the hybrid asymptotic-numerical value $q=\sigma E^{-1/2}-k^2\ell^{-2}$ agree so well that for the purpose of plotting  the graphs in figures~\ref{figk}($a$--$d$), they are essentially the same.  The weak dependence of $k$ and $\alpha$ on $E$ evident from their values itemised in each respective figure  proves to be a delicate issue that we discuss in \S\ref{side-wall}, but particularly \S\ref{side-wall-asym}.

\section{The transient evolution of the QG-azimuthal angular velocity \label{QG-trans}}

We consider the evolution of $\bar{\mathring \omega}(r,\tau)$, which solves 
\bme
\label{QG-eq-new}
\be
\pd{\bar{\mathring \omega}}{\tau}\,= \,\ell^2\biggl(\pd{^2 {\bar{\mathring \omega}}}{r^2}\,+\frac{3}{r}\pd{\bar{\mathring \omega}}{r}\biggr)\,=\,\dfrac{\ell^2}{r^3}\pd{\Gamma}{r}\,,\qquad\qquad 
\Gamma\,=\,r^3\pd{\bar{\mathring \omega}}r
\ee
\eme
(see (\ref{QG-eq-b})), on the $\tau=O(1)$ time-scale. Here $\bar{\mathring \omega}(r,\tau)$ defines the actual QG angular velocity
\be
\label{sol-tau-new}
\bar{\breve \omega}(r,t)\,=\,\kappa\,\bar{\mathring \omega}(r,\tau)\,\exp(-\sigma E^{1/2}t)
\ee
(see (\ref{sol-tau}$a$)), while $\Gamma$ provides a measure of the total viscous couple on cylinders of fluid radius $r$. We need to solve (\ref{QG-eq-new}) subject to the ($\tau=0$) initial condition
\be
\qquad{\bar{\mathring \omega}}\,=\,1  \qquad\qquad\quad \mbox{on} \qquad 0\,<\,r\,<\ell\,,
\label{QG-ic}
\ee
determined by the spin-down solution (\ref{simple-spin-down}$a$), and the boundary conditions
\be
\label{QG-bc}
\dfrac{1}{\ell^2}\Gamma \,=\, \ell\pd{\bar{\mathring \omega}}{r} \,= \left\{\begin{array}{lll}
0 \qquad&  {\rm at} \quad & r=0\,,\\[0.2em]
 {\mathring \alpha} \,{\bar{\mathring \omega}} \qquad &{\rm at} \quad & r=\ell\,.\end{array}\right.
\ee
With ${\mathring \alpha}=\ell\alpha>0$ the outer boundary condition $\Gamma(\ell,\tau)=\ell^2{\mathring \alpha}{\bar{\mathring \omega}}(\ell,\tau)$ (see (\ref{kalpha}$a$,$b$)) says that azimuthal motion at the outer boundary leads to a couple that accelerates rather than brake the total rotational angular momentum $r^3{\bar{\mathring \omega}}$ on cylinders radius $r$.

\cite{PH97} in their discussion of compressible Stewartson layers provided the complete transient solution of a diffusion equation similar to (\ref{QG-eq-new}). For our purposes it is sufficient to note that, as $\tau\to \infty$, the asymptotic solution is given by (\ref{QG-decay}). It is simply the first term of a modal expansion
\be
\label{QG-me}
\bar{\mathring \omega}(r,\tau)\,=\,A_0{\bar {\mathring \omega}}_0(r)\exp\bigl(k^2\tau\bigr)\,+\,\sum_{n=1}^\infty A_n{\bar {\mathring \omega}}_n(r)\exp\bigl(-k_n^2\tau\bigr)\,,
\ee
where ${\bar {\mathring \omega}}_0(r)$ defined by (\ref{Bessel-sol}) solves (\ref{Bessel-eq}). Each ${\bar {\mathring \omega}}_n(r)$ ($n>0$) satisfies the boundary conditions (\ref{QG-bc}) and solves (\ref{Bessel-eq}) with $k^2$ replaced by appropriate eigenvalues $-k_n^2$. It is easy to show that the corresponding eigenfunctions ${\bar {\mathring \omega}}_n(r)$ are orthogonal:
\be
\label{QG-orth}
\int_0^\ell{\bar {\mathring \omega}}_n{\bar{\mathring \omega}}_m r^3 \dR r\,=\,0   \qquad\qquad    m \not= n\,.
\ee
The initial condition $\bar{\mathring \omega}(r,0)=1$ thus determines
\be
\label{QG-asym}
\int_0^\ell {\bar {\mathring \omega}} r^3\,\dR r\,=\,A_0\int_0^\ell {\bar {\mathring \omega}}^{\,2} r^3\,\dR r\,,
\ee
which on substitution of ${\bar{\mathring \omega}}_0(r)=2(k r/\ell)^{-1}\,\IR_1(k r/\ell)$ (see (\ref{Bessel-sol})) gives (\ref{Bessel-A0}).

The solution corresponding to (\ref{QG-me}) for the \cite{GH63} problem with the rigid outer wall, at which ${\bar {\mathring \omega}}(\ell,\tau)=0$ is
\bse
\label{QG-me-GH}
\be
\bar{\mathring \omega}(r,\tau)\,=\sum_{n=1}^\infty 2B_n(j_nr/\ell)^{-1}\JR_1(j_nr/\ell)\exp\bigl(-j_n^2\tau\bigr)
\ee 
\citep[cf.][eq.~(19)]{PH97}, where $j_n(>0)$ are the zeros of the Bessel function $\JR_1$. The slowest decaying mode $n=1$ has
\be
j_1\,=\, 3.8317\cdots\,,
\ee
\ese 
while its amplitude $B_1$ is determined in the same way as $A_0$ above and is given in (\ref{ms-as-sol-GH}).

\subsection{Numerical results for the case $\ell=1$, $E=10^{-4}$\label{QG-num-res}}

We solved the reduced initial value problem (\ref{QG-eq-new}), (\ref{QG-ic}) and (\ref{QG-bc}) for ${\bar {\mathring \omega}}(r,\tau)$ numerically and compared the results with the $z$-average $\langle{\breve \omega}_\tDNS\rangle(r,\tau)$ obtained from the DNS solution ${\breve \omega}_\tDNS(r,z,t)$ of the entire problem (\ref{mom-vort})--(\ref{bc}) for ${\breve \omega}(r,z,t)$. To make that comparison we need to build on our analytic predictions. Firstly, the analysis of appendix~\ref{infinite-layer} indicates that the amplitude of ${\breve \omega}$ is modified by the factor $\kappa$ (see (\ref{simple-spin-down}$a$,$b$), also (\ref{kappa-mu}$a$)) on the spin-down time $t=O\bigl(E^{-1/2}\bigr)$, equivalently $\tau=O\bigl(E^{1/2}\bigr)$. Accordingly, we account for the spin-down decay rate and that amplitude modification in our analytic representation of the QG-mainstream angular velocity ${\bar {\breve \omega}}(r,t)$ in (\ref{sol-tau-new}). Secondly, our asymptotic results predict ${\bar {\breve \omega}}=\mu^{-1}\langle{\breve \omega}\rangle$ (see (\ref{MS-flux}$a$,$b$), also (\ref{kappa-mu}$b$)), and so we define the corresponding DNS value by
\be
\label{omega-DNS}
{\bar{\breve \omega}}_{\tDNS}(r,t)\,\equiv \,\mu^{-1}\langle {\breve \omega}_{\tDNS}\rangle(r,t)
\ee
as in (\ref{omega-ENS}). Thirdly, in view of the definition ${\mathring \omega}=\kappa^{-1}{\breve \omega}\exp\bigl(\sigma E^{1/2}t\bigr)$ (see (\ref{sol-tau})), we also define
\bse
\label{bar-av}
\be
{\bar{\mathring \omega}}_{\tDNS}(r,\tau)\,\equiv \,(\mu\kappa)^{-1}\langle{\breve \omega}_{\tDNS}\rangle(r,t)\exp(\sigma E^{1/2}t)\qquad\qquad\mbox{for}\qquad \tau=O(1)\,,
\ee
where
\be
(\mu\kappa)^{-1}\,=\,1\,+\,\tfrac14 E^{1/2}\,+\,O(E)
\ee
\ese
(see (\ref{kappa-mu}$a$,$b$)). 
\begin{figure}
\centerline{}
\vskip 3mm
\centerline{
\includegraphics*[width=0.8 \textwidth]{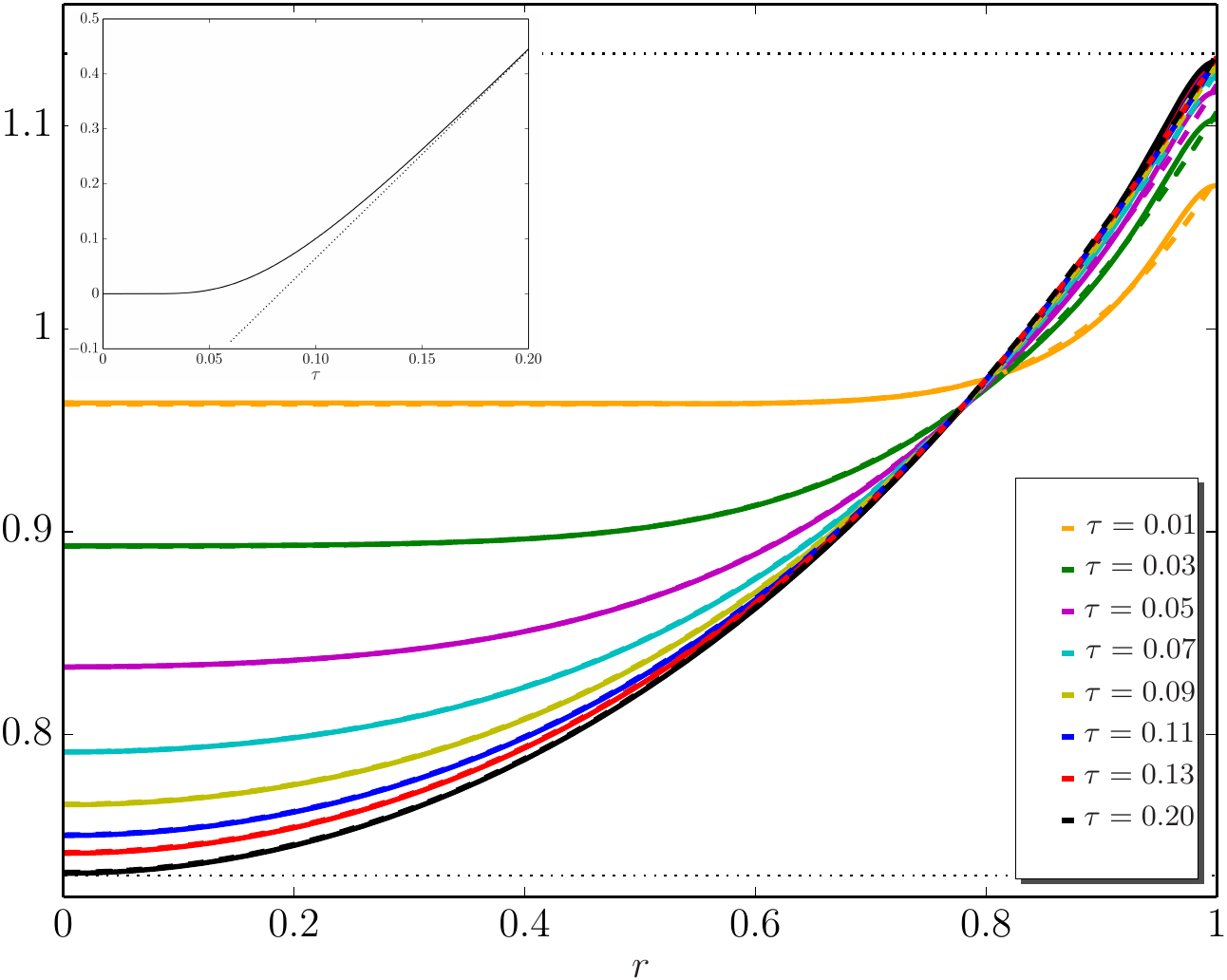}
}
\caption{(Colour online) Profiles of ${\bar{\mathring \omega}}_\tDNS(r,\tau)\exp\bigl(-k^2\tau\bigr)$ (continuous lines) for the case  $E=10^{-4}$, $\ell=1$, for which $k\doteqdot 1.9457$, ${\mathring \alpha}\doteqdot 0.8254$, $\sigma\doteqdot 1.0075$ (see data on figure~\ref{figk}($b$)), at various times $\tau$, together with the corresponding profiles of ${\bar{\mathring \omega}}(r,\tau)\exp\bigl(-k^2\tau\bigr)$ (dashed lines). The asymptotes ${\bar {\mathring \omega}}(0,\tau)\exp\bigl(-k^2\tau\bigr)\downarrow A_0\doteqdot 0.7307$ and ${\bar {\mathring \omega}}(\ell,\tau)\exp\bigl(-k^2\tau\bigr)$ $\uparrow A_\ell\doteqdot 1.1355$ as $\tau\to\infty$ are indicated by the dotted lines. ({\itshape{Inset}}) The left-hand end-point value $\log\bigl[\,{\bar{\mathring \omega}}(0,\tau)\bigr]$ plotted versus $\tau$, together with its large $\tau$ asymptote $(\log A_0)+k^2(\log\,\eR)\tau$.}
\label{transient}
\end{figure}

We illustrate the transient development of ${\bar {\mathring \omega}}(r,\tau)$ and ${\bar{\mathring \omega}}_{\tDNS}(r,\tau)$ for aspect ratio $\ell=1$ and $E=10^{-4}$ in figure~\ref{transient} with other data itemised in the caption. Since (\ref{QG-me}) determines the late time behaviour
\be
\label{bar-trans}
{\bar {\mathring \omega}}(r,\tau)\,\approx\,A_0\,{\bar {\mathring \omega}}_0(r)\,\exp\bigl(k^2\tau\bigr) \qquad\qquad\mbox{as}\qquad \tau\to\infty\,,
\ee
we remove this exponential growth and plot ${\bar {\mathring \omega}}(r,\tau)\exp\bigl(-k^2\tau\bigr)$ and ${\bar{\mathring \omega}}_{\tDNS}(r,\tau)\exp\bigl(-k^2\tau\bigr)$ instead at various times $\tau$. The $O(E^{1/2})$ correction to ${\bar{\mathring \omega}}_{\tDNS}$ (see (\ref{bar-av}$a$)) that ensues through the factor $(\mu\kappa)^{-1}$ (see (\ref{bar-av}$b$)) leads to very good agreement between the respective curves at each $\tau$. Indeed, they are largely indistinguishable except in  the $E^{1/3}$--sidewall-layer, where ${\bar{\mathring \omega}}_{\tDNS}$ needs to adjust to meet the boundary condition $\partial{\bar {\mathring \omega}}_{\tDNS}\big/\partial r=0$ at $r=\ell\,(=1)$. Similar discrepencies are visible in figure~\ref{figk}($b$).

 As noted in the early time asymptotics of the following \S\ref{QG-series} the left-hand (or centre) value  ${\bar {\mathring \omega}}(0,\tau)$ remains at unity until it eventually increases in response to diffusion across the domain driven by the right-hand (or outer) boundary condition. Eventually ${\bar {\mathring \omega}}(0,\tau)$ grows exponentially (see (\ref{bar-trans})). To illustrate this behaviour, we plot $\log\bigl[\,{\bar{\mathring \omega}}(0,\tau)\bigr]$ versus $\tau$ together with its large $\tau$ asymptote (recall that ${\bar {\mathring \omega}}_0(0)=1$) in figure~\ref{transient}({\itshape{Inset}}). It suggests a transition between small and large $\tau$ behaviour over roughly the range $0.05\lessapprox \tau \lessapprox 0.1$. This view is supported by figure~\ref{transient} itself, in which ${\bar {\mathring \omega}}(r,\tau)\exp\bigl(-k^2\tau\bigr)$ begins with the value unity at $\tau=0$. As time proceeds the flatness of the profiles near $r=0$ for $\tau=0.01$ and $0.03$ has largely disappeared by $\tau=0.11$. Subsequently the left ($r=0$) and right  ($r=\ell$) end point values decrease and increase (respectively) monotonically approaching the respective values $A_0$ and $A_\ell$ (see (\ref{left-right-end-values})) as $\tau\to\infty$. 

\subsection{A short time ($\tau\ll 1$) series expansion \label{QG-series}}

The initial condition ${\bar{\mathring \omega}}=1$ (see (\ref{QG-ic})) satisfies the governing equation (\ref{QG-eq-new}) and the boundary condition (\ref{QG-bc}) at $r=0$ but not at $r=\ell$. The discontinuity at $r=\ell$ is readily accommodated by the similarity solution 
\bme
\label{sing-eq}
\be
\ell\pd{\bar{\mathring \omega}}{r} \,= \,{\mathring \alpha} \erfc\zeta\,, \qquad\qquad\mbox{where} \qquad \zeta\,=\,\dfrac{1-(r/\ell)}{\sqrt{4\tau}}\,,
\ee
\eme
that solves $\partial {\bar{\mathring \omega}}\big/\partial \tau=\ell^2\partial^2{\bar{\mathring \omega}}\big/\partial r^2$ and meets the initial condition $\partial {\bar{\mathring \omega}}\big/\partial r(r,0)=0$ and the boundary conditions  $\ell\partial {\bar{\mathring \omega}}\big/\partial r (\ell,\tau)={\mathring \alpha}$ and $\ell\partial {\bar{\mathring \omega}}\big/\partial r \to 0$ as $\zeta\to\,\infty$. Integration with respect to $r$ yields
\bme
\label{sing-sol}
\be
{\bar{\mathring \omega}}\,=\,1\,+\,{\mathring \alpha}\sqrt{4\tau}\,\varpi_1(\zeta) \qquad\qquad\mbox{with} \qquad\varpi_1^{\,\prime}(\zeta)\,=\,-\,\erfc\zeta
\ee
giving
\be
\se
\varpi_1(\zeta)\,=\,\int_\zeta^{\infty}\erfc\zeta\,\,\dR\zeta\,=\,\dfrac{1}{\sqrt\pi}\exp\bigl(-\zeta^2\bigr)\,-\,\zeta\erfc\zeta\,.
\ee
\eme

\begin{figure}
\centerline{}
\vskip 3mm
\centerline{
\includegraphics*[width=1.0 \textwidth]{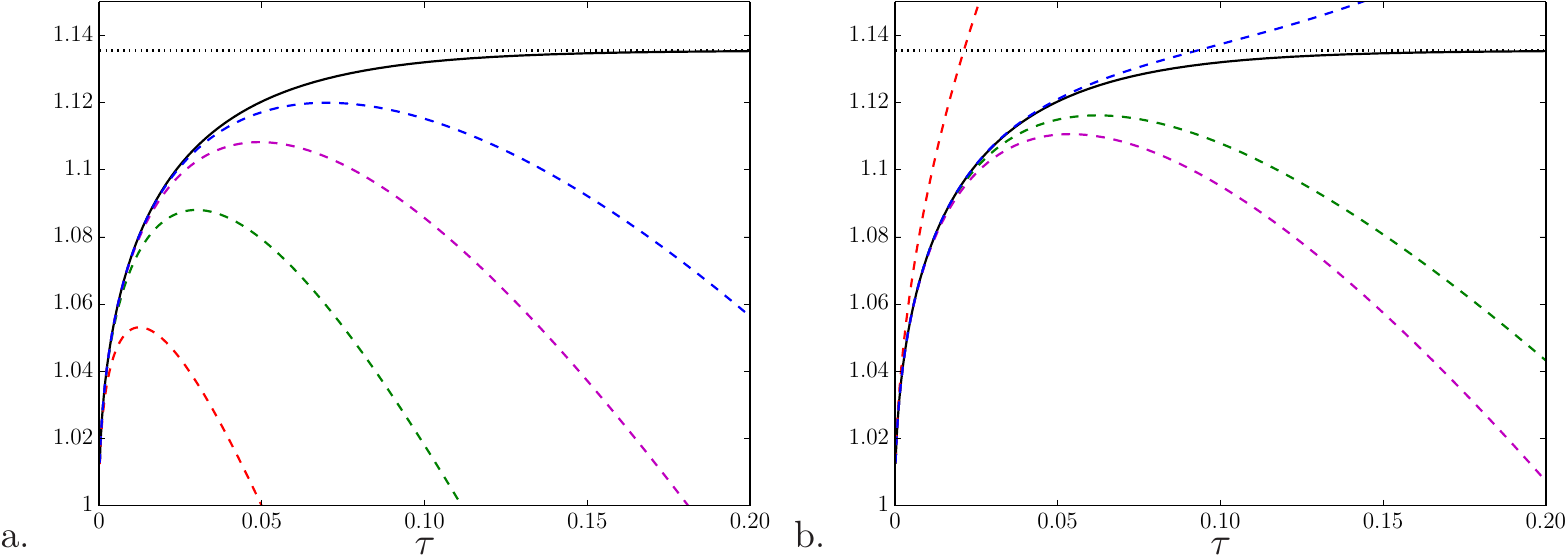}
}
\caption{(Colour online) The right-hand end-point value ${\bar{\mathring \omega}}(\ell,\tau)\exp\bigl(-k^2\tau\bigr)$ (solid line) plotted versus $\tau$ for the case illustrated in figure~\ref{transient}. ($a$) The dashed curves correspond to ${\bar{\mathring \omega}}(\ell,\tau)$ given by the series expansion (\ref{sim-series-ell}) truncated at various orders: $n=1$, Red; $2$, Green; $3$, Magenta; $4$, Blue (see (\ref{sim-series})). ($b$) As in ($a$) but the power series for $\exp\bigl(-k^2\tau\bigr)$ is employed and the product (\ref{sim-series-scaled}$b$) again approximated at the various orders. We also show the asymptote ${\bar {\mathring \omega}}(\ell,\tau)\exp\bigl(-k^2\tau\bigr)$ $\uparrow A_\ell\doteqdot 1.1355$ as $\tau\to\infty$ (dotted line).} 
\label{transient_end}
\end{figure}

The similarity solution (\ref{sing-sol}) is the first term of an asymptotic power series solution 
\be
\label{sim-series}
{\bar{\mathring \omega}}\,=\,1\,+\,{\mathring \alpha}\sum_{n=1}^\infty (4\tau)^{n/2} \varpi_n(\zeta)
\qquad\qquad \mbox{valid on}  \qquad\quad 0<\zeta< (4\tau)^{-1/2}
\ee
for $\tau\ll 1$. We outline the solution built around (\ref{sim-series-n})-(\ref{sim-series-fg}) in appendix~\ref{QG-pow-ser}. 

As a diagnostic for comparison with our direct numerical solution of (\ref{QG-eq-new})-(\ref{QG-bc}) we note that the end point value (\ref{sim-series-ell_A}) at $r=\ell$ determined by the terms (\ref{sim-series-nm}) and (\ref{sim-form-ser-fg-1})-(\ref{sim-form-ser-fg-4}), up to $n=4$, is
\begin{align}
{\bar {\mathring \omega}}(\ell,\tau)\,\approx 1\,&+\,\dfrac{2{\mathring \alpha}}{\sqrt \pi}\tau^{1/2}\,+\,{\mathring \alpha}\biggl(\dfrac32\,+\,{\mathring \alpha}\biggr)\tau\,+\,\dfrac{{\mathring \alpha}}{\sqrt\pi}\biggl(\dfrac52+4{\mathring \alpha}+\dfrac43{\mathring \alpha}^2\biggr)\tau^{3/2}\nonumber\\
&+\,\dfrac{3{\mathring \alpha}}{4}\biggl(\dfrac{5}{4}\,+\,4{\mathring \alpha}\,+\,3{\mathring \alpha}^2\,+\,\dfrac{2}{3}{\mathring \alpha}^3\biggr)\tau^2\,+\,O(\tau^{5/2})\,.
\label{sim-series-ell}
\end{align}
To compare with the large $\tau$ solution ${\bar {\mathring \omega}}(r,\tau)={\bar {\omega}}(r)\exp(k^2\tau)$, we expand ${\bar {\mathring \omega}}(r,\tau)\exp(-k^2\tau)$ at both $r=0$ and $r=\ell$ to the same order of accuracy:
\bse
\label{sim-series-scaled}
\begin{align}
&{\bar {\mathring \omega}}(0,\tau)\exp(-k^2 \tau)\,\approx \,\exp(-k^2 \tau)\,\approx \,1\,-\,k^2 \tau+\tfrac12\,k^4\tau^2\,+\,O(\tau^{3})\,,\\
&{\bar {\mathring \omega}}(\ell,\tau)\exp(-k^2 \tau)\,\approx \,1\,+\,\dfrac{2{\mathring \alpha}}{\sqrt \pi}\tau^{1/2}\,+\biggl[{\mathring \alpha}\biggl(\dfrac32\,+\,{\mathring \alpha}\biggr)\,-\,k^2\biggr]\tau\nonumber\\
&\qquad+\,\dfrac{{\mathring \alpha}}{\sqrt\pi}\biggl(\dfrac52+4{\mathring \alpha}+\dfrac43{\mathring \alpha}^2\,-\,2 k^2\biggr)\tau^{3/2}\nonumber\\
&\qquad+\biggl[\dfrac{3{\mathring \alpha}}{4}\biggl(\dfrac{5}{4}\,+\,4{\mathring \alpha}\,+\,3{\mathring \alpha}^2\,+\,\dfrac{2}{3}{\mathring \alpha}^3\biggr)-\,k^2{\mathring \alpha}\biggl(\dfrac32\,+\,{\mathring \alpha}\biggr)\,+\,\dfrac12\,k^4\biggr]\tau^2\,+\,O(\tau^{5/2})\,.
\end{align}
\ese

Returning to results from the direct numerical solution of (\ref{QG-eq-new})-(\ref{QG-bc}), whereas in figure~\ref{transient}({\itshape{Inset}}) we plot the logarithm of the left-hand end-point value ${\bar{\mathring \omega}}(0,\tau)$, in figure~\ref{transient_end}($a$) we plot the right-hand end-point value ${\bar{\mathring \omega}}(\ell,\tau)\exp\bigl(-k^2\tau\bigr)$. We also show the approximations derived by using the power series representation of ${\bar{\mathring \omega}}(\ell,\tau)$  given by the series expansion (\ref{sim-series-ell}) truncated at various levels. The first $n=1$ term truncation (\ref{sing-sol}) identifies the $\tau^{1/2}$ singular behaviour near $\tau=0$ but only gives a good approximation for very small $\tau$. Further terms improve the $\tau$-range of usefulness considerably. In figure~\ref{transient_end}($b$), we employ the complete power series representation (\ref{sim-series-scaled}$b$) of ${\bar{\mathring \omega}}(\ell,\tau)\exp\bigl(-k^2\tau\bigr)$, which interestingly for even $n$ (i.e., $n=2$, $4$) improves the approximation, exhibiting a longer $\tau$-range of usefulness at each level of truncation. The improvement is not apparent for odd $n$ (i.e., $n=1$, $3$). It is worth noting that the best ($n=4$) truncation provides a reasonable approximation up until $\tau\sim 0.1$, by which time the solution portrayed in figure~\ref{transient} is close to its final asymptotic $\tau\to\infty$ form. It must be appreciated that the power series solution (\ref{sim-series-ell_A}) described in appendix~\ref{QG-pow-ser} is truly asymptotic, not least because it takes no account of the left-hand ($r=0$) boundary condition.

\section{The $E^{1/3}$--side-wall shear-layer \label{side-wall}}

We pointed out in \S\ref{GH63-problem} that $E^{1/3}$--shear-layer adjacent to the outer cylinder wall $r=\ell$ is established on the time scale $t\sim E^{-1/3}$ short compared to the spin-up time $t\sim E^{-1/2}$. Once established its longer time evolution is best investigated via the modulation amplitudes ${\mathring \omega}(r,z,\tau)$ and ${\mathring \psi}(r,z,\tau)$ of the spin-down solution, which satisfy the scaled governing equations (\ref{mom-vort-l-t}$a$,$b$). Our ultimate objective is to establish analytically the value of $\alpha$ in (\ref{kalpha}$c$), that determines the mainstream boundary condition $\dR {\bar \omega}/\dR r=\alpha {\bar \omega}$ at $r=\ell$. This relation is only meaningful for $t\gg E^{-1/3}$ by which time $\alpha$ is well defined and independent of $t$. So to determine $\alpha$ it is sufficient to consider the final time eigenfunction
\be
\label{ef}
\kappa^{-1}\bigl[\,{\breve \psi}\,,\,{\breve \omega}\,\bigr]\exp\bigl(q\ell^2\tau\bigr)\,=\,\bigl[\,\psi\,,\,\omega\,\bigr](r,z)\,.
\ee
(cf.~(\ref{ms-lt}$a$,$b$) equivalently (\ref{sol-tau}$a$,$b$) and (\ref{QG-decay})). Though formulated in this restricted sense for clarity, our shear layer analysis pertains to all $t\gg E^{-1/3}$ upon making the approximation $q\ell^2\tau=E^{1/2}t$ in (\ref{ef}).

We restrict attention to the shear-layer flow outside the Ekman layer and represent it as the sum of the QG-mainstream solution (\ref{MS-sol}$a$,$b$) and a shear-layer correction:
\bme
\label{sl-var}
\be
 \omega\,=\,{\bar \omega}(r)\,+\,\epsilon\,{\bar \omega}(\ell)V(\zeta,z)\,, \qquad\quad
 \psi\,=\,{\bar \psi}(r)\,(z-1)\,+\,\ell^2\,{\bar \omega}(\ell)\,\Psi(\zeta,z)\,,
\ee
where 
\be
\te
{\bar \psi}(r)\,=\,-\,\tfrac12\sigma r^2 {\bar \omega}(r)\,,\qquad\qquad \zeta\,=\,\epsilon^{-2}(r-\ell)\qquad \mbox{and} \qquad \epsilon\,=\,E^{1/6}
\ee
\eme
(see (\ref{Bessel-sol}) and (\ref{MS-sol-new}), also (\ref{psi-QG})). Here $\zeta$ is the stretched radial boundary layer coordinate and $\epsilon$ is the natural expansion parameter: $E^{1/3}=\epsilon^2$ and $E^{1/2}=\epsilon^3$. The boundary layer corrections $V$, $\Psi$ must tend to zero as $\zeta\to -\infty$ as well ensure that the boundary conditions (\ref{bc}$a$) at $r=\ell$ are met.

To extract the boundary layer equations correct to $O(\epsilon)$, it is sufficient to consider (\ref{mom-vort-l-t}$a$,$b$) with $\sigma\approx 1$ (so, e.g., (\ref{sl-var}$c$) becomes ${\bar \psi}(r)=-\tfrac12 r^2 {\bar \omega}(r)$). They determine
\bme
\label{sl-eq}
\be
-\,2\pd{\Psi}z\,=\,\pd{^2 V}{\zeta^2}\,+\,\epsilon V\,, \qquad\qquad 
2\pd{V}z\,=\,\pd{^2\,}{\zeta^2}\biggl(\,\pd{^2\Psi}{\zeta^2}\,+\epsilon \Psi\biggr).
\ee
\eme
To the same order of accuracy, the boundary conditions (\ref{bc}$a$) are
\bse
\label{sl-bc}
\begin{align}
\Psi\,=\,\tfrac12 (z-1)\,,\qquad 
\pd{^2\Psi}{\zeta^2}\,=\,0\,,\qquad
\pd{V}{\zeta}\,&=\,-\,\epsilon\,\alpha & \mbox{at} &&\zeta\,=&\,0\,,\\
\Psi\,\to\,0\,,\qquad\qquad V\,&\to\,0\,,& \mbox{as} &&\zeta\,\to&\,-\,\infty
\intertext{for $0<z<1$, where $\alpha= \dR {\bar\omega}/\dR r(\ell)\big/{\bar\omega}(\ell)$, while the Ekman jump condition at $z=0$  on the flow exterior to the Ekman layer and the symmetry conditions (\ref{bc}$c$) at $z=1$ are respectively}
\Psi\,&=\,\tfrac12\epsilon\, V&  \mbox{at}&& z\,=&\,0\,,\\
\qquad{\Psi}\,=\,\pd{^2\Psi}{z^2}\,=\,\pd{V}{z}\,&\,=\,\,0&  \mbox{at} &&z\,=&\,1\qquad
\end{align}
\ese
for $\zeta<0$. 

In view of the boundary conditions (\ref{sl-bc}$b$), we may introduce the new variable
\bme
\label{Theta}
\be
\Theta(\zeta,z)\,=\,\int_{-\infty}^\zeta V\,\dR\zeta
\qquad\qquad\Longrightarrow \qquad\qquad    V\,=\,\pd{\Theta}{\zeta}\,,
\ee
\eme
and integrate  (\ref{sl-eq}$b$). By this device, we may symmeterise the shear-layer equations (\ref{sl-eq}$a$,$b$) and express them in the compact complex form
\bme
\label{sl-eq-sym}
\se
\be
2\iR\,\pd{\Xi}z\,=\,\pd{\,}{\zeta}\biggl(\,\pd{^2\Xi}{\zeta^2}\,+\epsilon\Xi\biggr),
\ee
where
\be
\de
\Xi\,=\,\Theta\,+\,\iR\,\Psi\qquad\quad\mbox{giving}\quad\qquad
\Upsilon\,=\,V\,+\,\iR\,W \,=\,\pd{\Xi}{\zeta} \qquad \biggl(\,W\,=\,\pd{\Psi}{\zeta}\biggr)
\ee
\eme
which is also useful.

\subsection{Asymptotic approach \label{side-wall-asym}}

We show in \S\ref{0-order} that, at zeroth order, the solution for $V(\zeta,z)$ has no $z$-mean, or more precisely, $\langle V\rangle=O(\epsilon)$. The objective of the first order problem investigated in \S\ref{side-wall-QG} is to solve (\ref{sl-eq-mean}) for $\langle V\rangle$ which, in turn, determines our key quantity of interest $\alpha=-\epsilon^{-1}\dR \langle V\rangle\big/\dR \zeta\big|_{\zeta=0}$, namely the $z$-average of (\ref{sl-bc}$a$)$_3$.

\subsubsection{The zeroth order problem \label{0-order}}

On setting $\epsilon=0$, the solution of (\ref{sl-eq})--(\ref{sl-eq-sym}) is
\bme
\label{sl-sol-0}
\se
\begin{align}
\Xi(\zeta, z)\,=\,&\,-\,\dfrac{2}{\sqrt3 \pi}\sum_{n=1}^\infty\dfrac{(-1)^n}{n}E\bigl(k_n\zeta;\,-\,\pi/6\bigr)\exp\bigl[\iR n\pi(z-1)\bigr],\\
\Upsilon(\zeta, z)\,=\,&\,-\,\dfrac{2}{\sqrt3 \pi}\sum_{n=1}^\infty\dfrac{(-1)^nk_n}{n}E\bigl(k_n\zeta;\,\pi/6\bigr)\exp\bigl[\iR n\pi(z-1)\bigr],
\end{align}
where
\be
\de   
k_n\,=\,(2n\pi)^{1/3},
\qquad\qquad
E\bigl(k_n\zeta;\,\alpha\bigr)\,=\,\exp(\tfrac12 k_n\zeta)\cos\bigl[\bigl(\sqrt{3}\big/2\bigr)k_n\zeta+\alpha\bigr].
\ee
\eme
Verification of the essential relation $\Upsilon=\partial \Xi/\partial \zeta$ (see (\ref{sl-eq-sym}$c$)) follows from the property
\bme
\label{E-prop}
\be
\od{\,}{\zeta} E\bigl(k_n\zeta;\,\alpha\bigr)\,=\,k_nE\bigl(k_n\zeta;\,\alpha+\pi/3\bigr)
\qquad\Longrightarrow\qquad
\od{^3E}{\zeta^3}\,=\,-\,k_n^3E
\ee
\eme
also. Since $\Im\{\exp[\iR n\pi(z-1)]\}=0$ at $z=0$ and $1$, the boundary conditions (\ref{sl-bc}$c$), namely $\Psi(\zeta,0)=0$, at the bottom and (\ref{sl-bc}$d$) at the top are all obviously met. Another property of the solution on $z=0$, useful later in \S\ref{side-wall-QG} (see (\ref{sl-V-mean}$b$)), is
\be
\label{int-Xi}
\int_{-\infty}^\zeta \Theta(\zeta,0)\,\dR \zeta\,=\,\int_{-\infty}^\zeta \Xi(\zeta,0)\,\,\dR \zeta\,=\,-\,\dfrac{2}{\sqrt3 \pi}\sum_{n=1}^\infty\dfrac{1}{nk_n}E\bigl(k_n\zeta;\,-\,\pi/2\bigr),
\ee
where we have used  (\ref{E-prop}$a$).

On use of (\ref{E-prop}$b$) it is readily seen that $\Xi$, defined by (\ref{sl-sol-0}$a$), solves (\ref{sl-eq-sym}$a$). Furthermore, since 
\be
\label{E-prop-more}
\od{^2}{\zeta^2} E\bigl(k_n\zeta;\,-\,\pi/6\bigr)\biggl|_{\zeta=0}\,=\,k_n^2E\bigl(k_n\zeta;\,\pi/2\bigr)\biggl|_{\zeta=0}\,=\,0\,,
\ee
it is clear that the two boundary conditions $\partial^2 \Psi/\partial \zeta^2=\partial V/\partial \zeta=0$ at $\zeta=0$ are met, i.e.,    
\be
\label{E-prop-bc}
\pd{^2\Xi}{\zeta^2}(0,z)\,=\,\pd{\Upsilon}{\zeta}(0,z)\,=\,0\,.
\ee
Furthermore, (\ref{sl-sol-0}$a$,$b$) evaluated at $\zeta=0$ determines
\bme
\label{Xi-Gamma-0}
\se
\begin{align}
\Xi(0,z)\,=\,&\,-\,\dfrac{1}{\pi}\sum_{n=1}^\infty\dfrac{(-1)^n}{n}\exp\bigl(\iR n\pi(z-1)\bigr)=\,\dfrac{1}{\pi}\,\ln\bigl[1\,+\,\exp\bigl(\iR \pi(z-1)\bigr)\bigr]\nonumber\\
=\,&\,\pi^{-1}\,\ln\bigl[2\,\sin\bigl(\tfrac12 \pi z\bigr)\bigr]+\,\tfrac{1}{2}\iR(z-1)\\
=\,&\,\dfrac{1}{\pi}\ln\bigl(\pi z\bigr)+\,\dfrac{\iR}{2}(-1+z)\,+\,\sum_{n=1}^\infty\Xi_n z^{2n}\,,\\  
\pd{\Xi}{\zeta}(0,z)\,=\,\Upsilon(0,z)\,=\,&\,-\,\dfrac{2^{1/3}}{\pi^{2/3}}\sum_{n=1}^\infty\dfrac{\exp(\iR n\pi z)}{n^{2/3}}
\,=\,-\,\dfrac{2^{1/3}}{\pi^{2/3}}\LR\iR_{2/3}\bigl(\eR^{\iR\pi z}\bigr)\nonumber\\
=\,&\,-\,\dfrac{2^{1/3}\Gamma(1/3)\eR^{\iR\pi/6}}{\pi z^{1/3}}\,+\,\sum_{n=0}^\infty\Upsilon_n z^n\,,
\end{align}
in which we have introduced the Polylogarithm $\LR\iR_{2/3}(\eta)=\sum_{n=1}^\infty n^{-2/3}z^n$, and where 
\be
\de
\Xi_n\,=\,-\,\dfrac{\zeta(2n)}{2^{2n}n\pi}\,,
\qquad\qquad\qquad
\Upsilon_n=\,-\,\dfrac{2^{1/3}}{\pi^{2/3}}\zeta\bigl((2/3)-n\bigr)\,\dfrac{(\iR \pi)^n}{n!}\,,
\ee
\eme
in which  $\zeta(m)\,\bigl(=\,\sum_{n=1}^\infty n^{-m}$, for $m>1$ and by analytic continuation elsewhere$\bigr)$ is the Riemann-Zeta function. The result (\ref{Xi-Gamma-0}$a$) verifies that the boundary condition $\Psi(0,z)=\tfrac{1}{2}(z-1)$ is met, while in addition determining the result $\Theta(0,z)=\pi^{-1}\,\ln\bigl[2\,\sin\bigl(\tfrac12 \pi z\bigr)\bigr]$. The series representations (\ref{Xi-Gamma-0}$b$,$c$) follow on use of (http://dlmf.nist.gov/25.8.E8), (http://dlmf.nist.gov/25.12.E12) respectively \citep[here and below, we use this style of online reference to equations in][]{AS10} and are valid over the entire range $0<z\le 1$.

\subsubsection{The power series solution\label{0-ps}}

The power series forms (\ref{Xi-Gamma-0}$b$,$c$) of the solution at $\zeta=0$ identify clearly a singularity at $(\zeta, z)=(0,0)$. To resolve it, we reconstruct the full solution (\ref{sl-sol-0}$a$,$b$) by solving the shear-layer equation (\ref{sl-eq-sym}$a$) subject to the boundary conditions (\ref{E-prop-bc}) and (\ref{Xi-Gamma-0}$b$,$c$) at $\zeta=0$. Those suggest the expansion
\bse
\label{Xi-Gamma-sim}
\begin{align}
\Xi(\zeta,z)\,=\,&\biggl[\dfrac{1}{\pi}\ln(\pi z)\,-\,\dfrac{\iR}{2}\,+\,\YS(\Phi)\biggr]+\,z\biggl[\dfrac{\iR}{2}\,-\,\dfrac16\Phi^3\biggr] +\,\Xih(\Phi,z)\,,\\
\Upsilon(\zeta,z)\,=\,&\,z^{-1/3}\ZS(\Phi)\,-\,\dfrac12 z^{2/3}\Phi^2\, + \,\Upsilonh(\Phi,z)\,,
\end{align}
where $\Phi$ is the similarity variable
\be
\Phi\,=\,\zeta\big/z^{1/3}\,.
\ee
\ese 
Each of the three terms in (\ref{Xi-Gamma-sim}$a$) when grouped with the corresponding term in (\ref{Xi-Gamma-sim}$b$) solves  (\ref{sl-eq-sym}$a$). The first terms, the similarity solution, resolve the singularity. The second and the third terms, namely the sums 
\bse
\label{Xi-Gamma-N}
\begin{align}
\Xih_N(\Phi,z)\,=\,&\sum_{n=0}^N \Upsilon_n \YS_n^+(\Phi) z^{n+1/3}\,+\,\sum_{n=1}^{[N/2]} \Xi_n \YS_{2n}(\Phi) z^{2n}\,,\\
\Upsilonh_N(\Phi,z)\,=\,&\sum_{n=0}^N \Upsilon_n\YS_n(\Phi)z^n\,+\,\sum_{n=1}^{[N/2]}\Xi_n \YS_{2n}^-(\Phi) z^{2n-1/3}
\end{align}
\ese
($[N/2]$ denotes integer part of $N/2$) as $N\to\infty$, are simply polynomial expressions. They are rendered unique by the boundary conditions (\ref{E-prop-bc}) and (\ref{Xi-Gamma-0}$b$,$c$) at $\zeta=0$, which require
\bme
\label{Yn0}
\be
\te
\YS(0)\,=\,0\,,\qquad\qquad   \YS_n(0)\,=\,1\,,\qquad\qquad    \YS_n^\pm(0)\,=\,0\,.
\ee
\eme
It is our belief that, like the exponentials which they approximate, the series (\ref{Xi-Gamma-N}) have an infinite radius of convergence (in $\zeta$) and hence meet the boundary condition $\Xi\to 0$ as $\zeta \to -\infty$. 

The required generating polynomial
\bme
\label{Y-def}
\be
\YS_n(\Phi)\,=\,\sum_{k=0}^{n}\dfrac{(2\iR)^{k} n!}{(3k)!(n-k)!}\Phi^{3k} 
\ee
with $\YS_n(0)=1$ (see (\ref{Yn0}$b$)) solves
\be
\LS_n\YS_n\,=\,0\,,\qquad\quad\mbox{where}\qquad\quad
\LS_n\,\equiv\,\od{^3}{\Phi^3}\,+\,\dfrac23 \iR\biggl(\Phi\od{\,}{\Phi}\,-\,3n\biggr).
\ee
\eme
It determines
\bme
\label{Ypm-def}
\be
\YS_n^+(\Phi)\,=\,-\,\int_\Phi^0 \YS_n(\Phi)\,\dR\Phi\,,\qquad\quad\mbox{and}\qquad\quad
\YS_n^-(\Phi)\,=\,\od{\YS_n}{\Phi}\,,
\ee
which solve
\be
\se
\LS_{n\pm 1/3}\YS^{\pm}_n\,=\,0\,,
\ee
\eme
and have the property $\YS_n^\pm(0)=0$ (see (\ref{Yn0}$c$)).

The similarity form $\pi^{-1}\ln(\pi z)-\tfrac12 \iR+\YS(\Phi)$ solves  (\ref{sl-eq-sym}$a$) with $\epsilon=0$ when
\bme
\label{Y-sol}
\be
\LS_0\YS\,=\,\dfrac{2\iR}{\pi}\qquad\qquad\Longleftrightarrow\qquad\qquad
\od{^2\ZS}{\Phi^2}\,+\,\dfrac23 \iR\Phi\ZS\,=\,\dfrac{2\iR}{\pi}\,,
\ee
where the limits of integration in
\be
\se
\YS(\Phi)\,=\,-\,\int_\Phi^0 \ZS(\Phi)\,\dR\Phi
\ee
\eme
are chosen such that $\YS(0)=0$ (see (\ref{Yn0}$a$)). The solution of (\ref{Y-sol}$b$), which satisfies the remaining boundary conditions 
\bme
\label{Y-sol-extra}
\be
\dR \ZS/\dR \Phi (0) \,=\,0 \qquad\quad\mbox{and}\qquad\quad \ZS\to 0\quad \mbox{as}\quad \Phi\to -\infty\,,
\ee
\eme
is
\bme
\label{Xi-Psi-sim-Scorer-sol}
\be
\se
\ZS(\Phi)\,=\,-\,\sqrt3\Bigl(\dfrac23\Bigr)^{\!1/3}
\biggl[\eR^{\iR\pi/3}\HR\iR\biggl(\iR\Bigl(\dfrac23\Bigr)^{\!1/3}\Phi\biggr)
+\,\HR\iR\biggl(\Bigl(\dfrac23\Bigr)^{\!1/3}\eR^{-\iR\pi/6}\Phi\biggr)\biggr],
\ee
where
\be
\HR\iR(\zS)\,=\,\dfrac{1}{\pi}\int_0^\infty\exp\bigl(-\tfrac13 \tS^3\,+\,\zS\tS\bigr)\,\dR\tS\,,
\qquad\quad\mbox{with}\qquad\quad
\HR\iR(0)\,=\,\dfrac{\Gamma(1/3)}{3^{2/3}\pi}\,,
\ee
is the Scorer function (http://dlmf.nist.gov/9.12.E20), which satisfies
\be
\se
\HR\iR^{\,\prime\prime} \,-\,\zS\,\HR\iR\,=\,1/\pi\,,
\ee
\eme
where the prime denotes derivative. Together (\ref{Xi-Psi-sim-Scorer-sol}$a$,$c$) determine
\be
\label{Scorer-0}
\ZS(0)\,=\,-\,\dfrac{2^{1/3}\Gamma(1/3)}{\pi}\eR^{\iR\pi/6}\,,
\ee
as required by the boundary condition (\ref{Xi-Gamma-0}$c$).

It is worth reflecting at this point, that the governing equation (\ref{Y-sol}$b$) is essentially that given by \cite{DS07} eq.~(3.26$b$) in their interpretation of the point source similarity solution derived by  \cite{MS69}. However, our lowest order combined boundary conditions $\partial V/\partial \zeta=0$ and $\Psi=-\tfrac12$ on $\zeta=0$ break the symmetry of the classical \cite{S57} and \cite{MS69} solutions. So whereas, the \cite{MS69} solution only involves a single Scorer function \citep[see][eq.~(3.26$a$)]{DS07}, we need two (see (\ref{Xi-Psi-sim-Scorer-sol}$a$) so complicating matters. From a more general point of view, our solutions are valid on the semi-infinite interval $-\infty<\Phi\le 0$ and would diverge if extended to $0\le\Phi<\infty$. By contrast, \cite{MS69} considered similarity solutions valid on the infinite interval $-\infty<\Phi <\infty$. These have proved useful in other $E^{1/3}$ shear-layer contexts \citep[see, e.g.,][eqs.~(4.1)-(4.3)]{MDS16}.

Substitution of (\ref{Xi-Psi-sim-Scorer-sol}$a$) into the integral (\ref{Y-sol}$c$) determines 
\bse
\label{Y-sim-Scorer-sol}
\be
\YS(\Phi)\,=\,\sqrt3\biggl[\eR^{-\iR\pi/6}J\biggl(-\,\iR\Bigl(\dfrac23\Bigr)^{\!1/3}\Phi\biggr)+\,\eR^{\iR\pi/6}J\biggl(-\Bigl(\dfrac23\Bigr)^{\!1/3}\eR^{-\iR\pi/6}\Phi\biggr)\biggr],
\ee
in which 
\be
J(\zS)\,=\,\int_0^\zS \HR\iR(-\breve\zS)\,\dR\breve\zS\,.
\ee
\ese
For large negative $\Phi$ the asymptotic result (http://dlmf.nist.gov/9.12.E31) determines
\be
\label{Y-sim-Scorer-asym}
\YS(\Phi)\,\approx\,\dfrac{1}{\pi}\bigl[3\ln(-\Phi)\,+\,2\gamma\,+\ln2\bigr]\,+\,\dfrac{1}{\sqrt{3}}\,+\,\dfrac{\iR}{2}\,+\dfrac{3\iR}{\pi(-\Phi)^3}\qquad\qquad (-\Phi\gg 1)\,,
\ee
where $\gamma=0.57721566\cdots$ is Euler's constant. 

Values close to the bottom boundary $z=0$ are important. On substitution (\ref{Y-sim-Scorer-asym}) into (\ref{Xi-Gamma-sim}$a$), the real and imaginary parts determine 
\bme
\label{Y-Z-asym}
\be
\Theta\big|_{z\downarrow 0}\,\approx\,\dfrac{1}{\pi}\bigl[3\ln(-\zeta)\,+\,2\gamma\,+\ln(2\pi)\bigr]\,+\,\dfrac{1}{\sqrt{3}}\,,\qquad\qquad \Psi\big|_{z\downarrow 0}\,\approx\,0\,
\ee
(note that $(-\Phi)^3\big|_{z\downarrow 0}= 0$, provided $\zeta\not=0$), from which we also deduce that
\be
\se
V\big|_{z\downarrow 0}\,=\,\pd{\Theta}{\zeta}\bigg|_{z\downarrow 0}\,\approx\,\dfrac{3}{\pi\zeta}
\qquad\qquad\mbox{for}\qquad    1\gg -\zeta \gg z^{1/3}\,.
\ee
Also of interest is the integral (\ref{int-Xi}), which evidently vanishes at $\zeta=0$. Its approach to zero is determined by (\ref{Y-Z-asym}$a$), which determines
\be
\int_{-\infty}^\zeta \Theta\big|_{z\downarrow 0}\,\dR \zeta\,\approx\,\dfrac{3}{\pi}\,\zeta\ln(-\zeta)
\qquad\qquad\mbox{for}\qquad    1\gg -\zeta \gg z^{1/3}\,,
\ee
\eme
needed for our understanding of ${\bar V}(\zeta)$ (see (\ref{sl-V-mean}$b$)) in the next \S\ref{side-wall-QG}). Of course, the limit $z\downarrow 0$ must be interpreted in the sense of our solution being valid outside the Ekman layer. As the Ekman layer width is $O(E^{1/2})$, we expect (\ref{Y-Z-asym}$c$) to be valid for $-\zeta \gg E^{1/6}$ (equivalently $\ell-r\gg E^{1/2}$).

In appendix~\ref{similarity-4}, we evaluate $\Psi$ and $V$ (see (\ref{Xi-Gamma-sim}$a$,$b$)) using the power series (\ref{Xi-Gamma-N}$a$,$b$) (see (\ref{Psi-V-origin-more}$a$,$b$)) and compare our results with those from the Fourier series (\ref{sl-sol-0}$a$,$b$) in figure~\ref{Fourier-similarity} below. Though the Fourier series needs a very large number of terms ($10^5$) to give an accurate representation near $(\zeta,z)=(0,0)$, only a very few terms ($N=4$) of the power series are needed to give a very accurate solution there. Indeed it is remarkable how good the power series solution is throughout the domain illustrated in figure~\ref{Fourier-similarity}. On increasing $-\zeta$ serious discrepancies on the entire range $0<z\le 1$ only begin to emerge at about $\zeta=-2.0$, by which distance the values of both $\Psi$ and $V$ are small and relatively close to zero. We find this high degree of accuracy from such a low truncation quite striking. Indeed our belief is that the differences visible are due to the truncation and that the Fourier series and power series solutions coincide in the limit $N\to \infty$.

The streamfunction $\Psi$ in figure~\ref{Fourier-similarity}($a$) shows uniform blowing from the right-hand boundary $0<z<1$, where $\Psi=\tfrac12(z-1)$ (see (\ref{sl-bc}$a$)$_1$) which, because $\Psi=0$ on $z=0$ (see (\ref{sl-bc}$b$)), is all returned into the bottom right-hand corner $(\zeta,z)=(0,0)$ so forming half an eddy. This is followed by a reverse flow eddy also spawned at $(\zeta,z)=(0,0)$. Further eddies follow but they are essentially too small to be visible. The $V$--contours in figure~\ref{Fourier-similarity}($b$) reflect the response of $V$ to the Coriolis acceleration caused by the radial flow, proportional to $-\partial \Psi/\partial z$. The singularity  in the corner identified by both $V\approx 3/(\pi\zeta)$ on $z=0$ (see (\ref{Y-Z-asym}$c$)) and  $V\propto z^{-1/3}$ on $\zeta=0$ (see (\ref{Xi-Gamma-0}$c$) and the red dashed curve, restricted to small $z$, in figure~\ref{Psi_z=0}($b$) below) gives the distinctive corner structure. The vertical nature of the $V$--contours close to the boundary at moderate $-\zeta\,(\approx 2.0)$ over a thin $z$--width is suggestive of an Ekman layer. That is an illusion, as in our asymptotics the Ekman layer has zero width. Rather it reflects how small $z$ needs to be to achieve the large $-\Phi$ asymptotic behaviour. Though not visible in the figure, this feature continues up to $\zeta=0$, where it is truly a characteristic of the similarity solution $V=z^{-1/3}\Re\{\ZS(\Phi)\}$, which determines the $V\approx 3/(\pi\zeta)$ behaviour alluded to.

\subsubsection{The first order (QG) problem \label{side-wall-QG}}

As the zeroth order solution $V(\zeta,z)=\Re\{\Upsilon(\zeta,z)\}$ defined by (\ref{sl-sol-0}$b$) has no mean part ($\langle V\rangle=0$), we need to consider the next $O(\epsilon)$ problem. On use of the boundary conditions (\ref{sl-bc}$c$,$d$), the $z$-mean of (\ref{sl-eq}$a$) determines
\bme
\label{sl-eq-mean}
\be
V(\zeta,0)\,=\,\od{^2{\bar V}}{\zeta^2}\,+\,\epsilon {\bar V}\,, \qquad\quad\mbox{where}\quad \qquad \langle V\rangle\,=\,\epsilon {\bar V}\,.
\ee
\eme
Since our shear-layer solution only concerns the flow outside the Ekman layer, we use the bar-notation for consistency with our $\bar{\omega}$ notation for the mainstream QG-solution. As $\langle V\rangle=O(\epsilon)$, the additional factor $\epsilon$ is included in (\ref{sl-eq-mean}$b$) for convenience. So ignoring the term $\epsilon {\bar V}$ in (\ref{sl-eq-mean}) on the basis that it is $O(\epsilon)$, integration with respect to $\zeta$, noting the definition (\ref{Theta}$a$), determines
\bme
\label{sl-V-mean}
\be
\od{\bar V}{\zeta}(\zeta)\,=\,\Theta(\zeta,0)
\qquad\qquad\Longrightarrow\qquad\qquad
{\bar V}(\zeta)\,=\,\int_{-\infty}^\zeta \Theta(\zeta,0)\,\dR \zeta
\ee
\eme
on further integration.

Our objective is to determine the value $\alpha=-\dR{\bar V}/\dR\zeta\big|_{z=0}$ postulated in the boundary condition (\ref{sl-bc}$a$)$_3$. Unfortunately (\ref{Y-Z-asym}$a$) indicates that $\dR{\bar V}/\dR\zeta(\zeta)=\Theta(\zeta,0)\propto \ln(-\zeta)$ 
\clearpage

\begin{figure}
\centerline{
\includegraphics*[width=1.0 \textwidth]{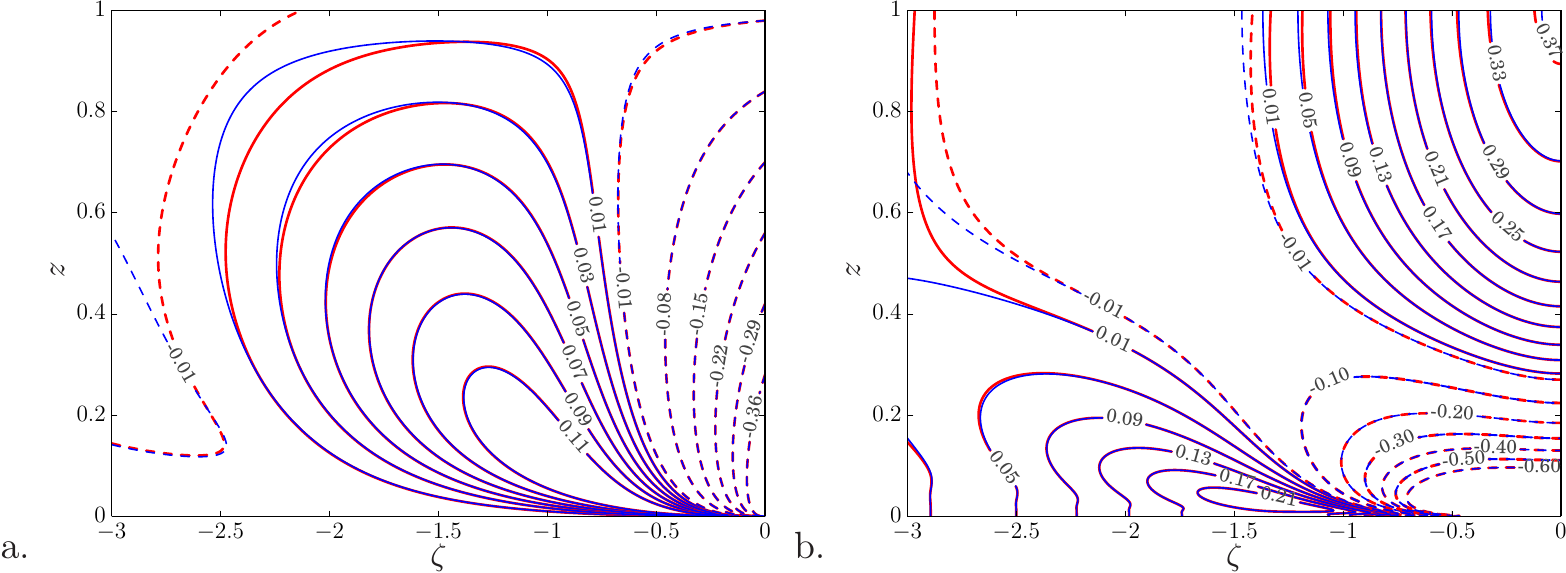}
}
\caption{(Colour online) Contours of ($a$) streamfunction $\Psi$ and ($b$) azimuthal velocity $V$, derived from the complex forms (\ref{sl-eq-sym}$b$,$c$), in the $\zeta$--$z$ plane. The Fourier series solution (\ref{sl-sol-0}) is red and the similarity solution (\ref{Xi-Gamma-sim}) with the power series (\ref{Xi-Gamma-N}) truncated at $N=4$ (see (\ref{Psi-V-origin-more})) is blue. Negative values are identified by dashed lines.} 
\label{Fourier-similarity}
\end{figure}

\begin{figure}
\centerline{
\includegraphics*[width=1.0 \textwidth]{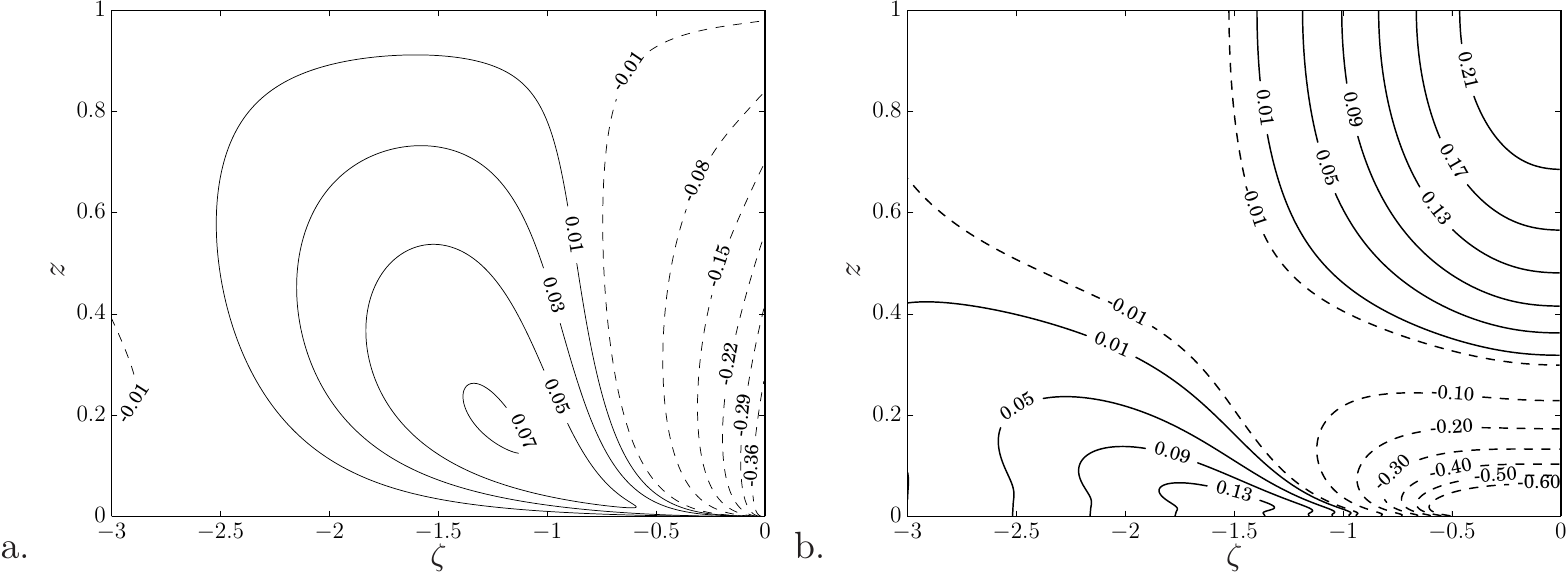}
}
\caption{As in figure~\ref{Fourier-similarity} for the case $E=10^{-5}$ ($\epsilon\approx 0.15$) for which the left-hand boundary $\zeta=-3.0$ corresponds to $r=0.93\cdots$, when $\ell=1$: Contours of ANS fluctuating ($a$)~streamfunction $\Psi^{\,\prime}_\tANS$ and ($b$)~azimuthal velocity $V^{\,\prime}_{\!\tANS}$ (see (\ref{ANS-prime}$a$,$b$)).}
\label{ANS-fluct}
\end{figure}

\begin{figure}
\centerline{}
\centerline{
\includegraphics*[width=1.0 \textwidth]{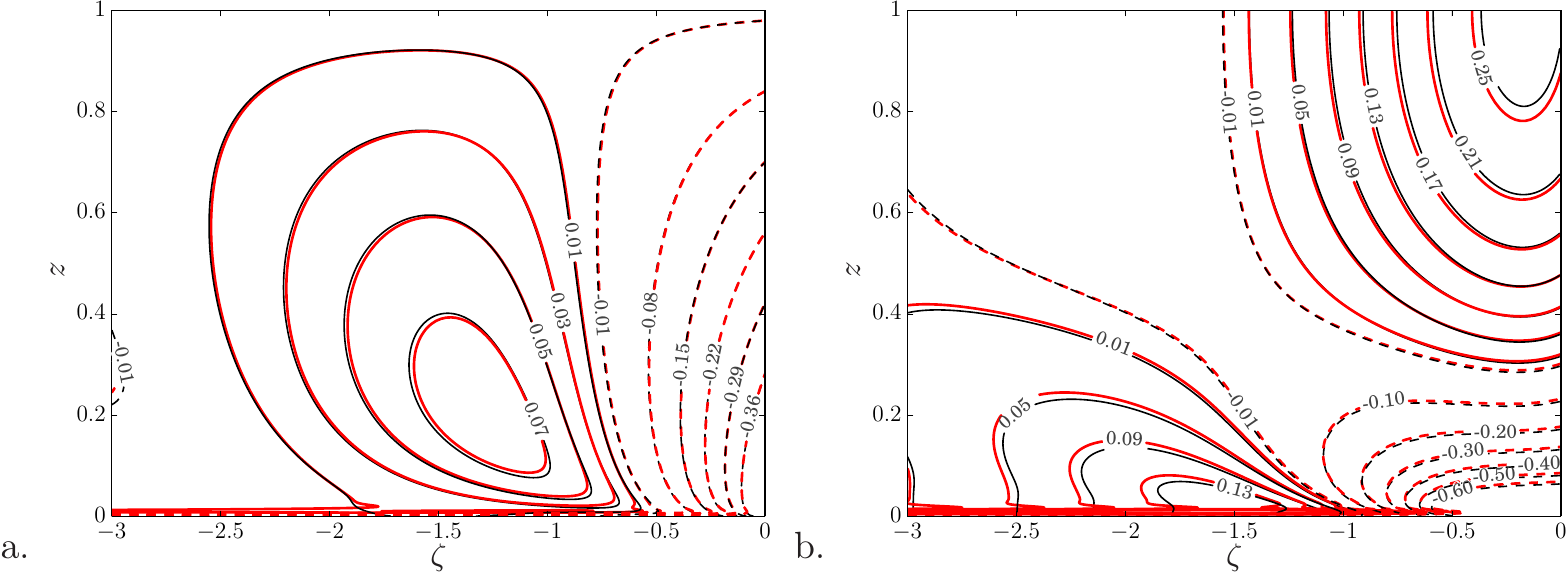}
}
\caption{(Colour online) As in figure~\ref{ANS-fluct} ($E=10^{-5}$, $\ell=1$): Red ENS-contours, ($a$) $\Psi_\tENS$ and ($b$) $V_\tENS$ (see (\ref{sl-var-ENS}$a$,$b$)); black ANS-contours, ($a$) $\Psi_\tANS$ and ($b$) $V_{\!\tANS}$.}
\label{ANS-ENS}
\end{figure}
\clearpage

\noindent
as $\zeta\uparrow 0$, the divergence of which means that $\alpha$ is not defined. This failure of our zeroth order asympotics is limited to the corner region $\ell-r=O(E^{1/2})$, $z=O(E^{1/2})$, equivalently $-\zeta=O(\epsilon)$, $z=O(\epsilon^3)$. It identifies the $\zeta$--length $O(\epsilon)$, at which we must cut-off $\ln(-\zeta)$ to obtain $\ln(\epsilon)+O(1)$, and whence
\be
\label{sl-ln-E}
\alpha\,=\,\alpha_\tFS\,\approx\,-\,\Theta(-\epsilon,0)\,\approx\,-\,\dfrac{3}{\pi}\ln\epsilon\,+\,O(1)\,=\,\dfrac{1}{2\pi}\ln\bigl(E^{-1}\bigr)\,+\,O(1)\,,
\ee 
where the subscript FS identifies the Fourier series origin (\ref{sl-sol-0}) of this result. Recall that (\ref{sl-ln-E}) is obtained from the similarity solution (\ref{Y-sim-Scorer-sol}$a$), which is part of the power series solution (\ref{Xi-Gamma-sim}) believed to be equivalent to (\ref{sl-sol-0}). Henceforth, all results based on either (\ref{sl-sol-0}) or (\ref{Xi-Gamma-sim})  will be likewise distinguished.

\subsection{Hybrid asymptotic-numerical solutions \label{side-wall-hybrid}}

Though the result (\ref{sl-ln-E}) is asymptotically sound, it is impossible to determine the $O(1)$ contribution to $\alpha_\tFS$ without solving the problem for the $E^{1/2}\times E^{1/2}$ corner region, which we have not attempted. Indeed, as $\epsilon=E^{1/6}$ is our expansion parameter, it is quite clear that no agreement with the full ENS is to be expected. Not only is $\epsilon\approx 0.1468$ for $E=10^{-5}$ (as employed in figure~\ref{figk}($d$)) not small, certainly $\ln\epsilon$ itself is an $O(1)$ number like the error neglected. To bypass this difficulty, we solved the problem posed by (\ref{sl-eq}) and (\ref{sl-bc}) numerically at small but finite $\epsilon$. As a preliminary test of the approximation, we undertook the numerical solution with $\epsilon=0$. Despite the singularity at the corner $(\zeta,z)=(0,0)$, the numerical results elsewhere agreed well with the Fourier series results portrayed in figure~\ref{Fourier-similarity}. Certainly to graph plotting accuracy the contours of constant $\Psi$ and $V$ are indistinguishable and so have not been re-plotted.

Henceforth our results obtained from our blend of asymptotic theory and numerical simulations (ANS) will be labelled by the subscript ANS. Though our implementation of ANS, adopts the boundary condition (\ref{sl-bc}$a$)$_3$:
\bme
\label{sl-bc-ANS}
\be
\pd{V_\tANS}{\zeta}(0,z)\,=\,-\,\epsilon\alpha_\tANS\,,
\qquad\quad\mbox{the value}\qquad\quad
\alpha_\tANS\,=\,-\,\od{{\bar V}_{\!\!\tANS}}{\zeta}(0)
\ee
\eme
is part of the answer. Indeed this consideration highlights the complication that the $\epsilon=0$ results, portrayed in figure~\ref{Fourier-similarity}, do not determine $\alpha_\tANS$. Furthermore, despite appearances from our scalings, the value of $\alpha_\tANS$ is itself a function of $\epsilon=E^{1/6}$ (asymptotically $-\ln\epsilon$, see (\ref{sl-ln-E})).

The purpose of the ANS-problem is to retain all the leading order elements of the asymptotic FS-problem, but also include the important $O(\epsilon)$ terms. To this end we considered in detail the aforementioned case $E=10^{-5}$. Even though $\epsilon\approx 0.1468$ is only moderately small, the comparisons made in figures~\ref{Fourier-similarity}--\ref{ANS-ENS} are really encouraging; remember that the errors in the ANS-method are of order $\epsilon^2\approx 0.0215 $, i.e., $2\%$.

Since $V_\tFS$ (i.e., $V_{\!\tANS}$, when $E=0$) displayed in figure~\ref{Fourier-similarity}($b$) has no mean part, we plot the fluctuating part
\bse
\label{ANS-prime}
\be
V^{\,\prime}_\tANS(\zeta,z)\,=\,V_\tANS(\zeta,z)\,-\,\epsilon{\bar V}_{\!\!\tANS}(\zeta)
\ee
in figure~\ref{ANS-fluct}($b$). Their constant--$V^{\,\prime}_\tANS$ contours compare well with those for $V_\tFS$ in figure~\ref{Fourier-similarity}($b$), but their numerical values differ by amounts that increase as $-\zeta$ approaches zero. The notion of a fluctuating part of $\Psi_\tANS$ is less clear. So noting that $V^{\,\prime}_\tANS$ measures the departure from quasi-geostrophy, we choose to plot the quantity
\be
\Psi^{\,\prime}_\tANS(\zeta,z)\,=\,\Psi_\tANS(\zeta,z) + \epsilon V_\tANS(\zeta,0)\bigl[-\,\Psi_\tFS(\zeta,z)\,+\,\tfrac12(z-1)\bigr],
\ee
\ese
which purports to measure the $\Psi$-departure from QG, in figure~\ref{ANS-fluct}($a$). The contribution $\epsilon V_\tANS(\zeta,0)\tfrac12(z-1)$ is the QG-part stemming from the Ekman suction boundary condition $\Psi_\tANS(\zeta,0)=\frac12\epsilon V_\tANS(\zeta,0)$. Furthermore as the non-zero value of $\Psi_\tANS(0,0)$ alters the amount of fluid flux that the shear-layer carries, we accommodate that effect by incorporating the corresponding Fourier solution combination $-\Psi_\tFS(\zeta,z)+\tfrac12(z-1)$ scaled by the lower boundary velocity $V_\tANS(\zeta,0)$. The upshot is that $\Psi^{\,\prime}_\tANS$ defined by (\ref{ANS-prime}$b$) satisfies the boundary conditions $\Psi^{\,\prime}_\tANS(\zeta,0)=0$ and $\Psi^{\,\prime}_\tANS(0,z)=\tfrac12 (z-1)$, exactly like the $\epsilon=0$ Fourier series $\Psi_\tFS=\Im\{\Xi\}$ (see (\ref{sl-eq-sym}$b$)). Note too that $\Psi^{\,\prime}_\tANS\to\Psi_\tFS$ and $V^{\,\prime}_\tANS\to V_\tFS$ as $\epsilon\to 0$. The comparison of  $\Psi^{\,\prime}_\tANS$, $V^{\,\prime}_\tANS$ for non-zero $\epsilon$ ($E=10^{-5}$) in figures~\ref{ANS-fluct}($a$,$b$) with $\Psi_\tFS$, $V_\tFS$ in figures~\ref{Fourier-similarity}($a$,$b$) is topologically very good. However, quantitative agreement is moderate, which is to be expected as the expansion parameter $\epsilon\approx 0.1468$ is not really small. It should be emphasised that, whereas $V^{\,\prime}_{\!\tANS}$ is a natural quantity, $\Psi^{\,\prime}_{\!\tANS}$ is not, but rather has been constructed to provide an analogue to $V^{\,\prime}_{\!\tANS}$.

\begin{figure}
\centerline{}
\vskip 3mm
\centerline{
\includegraphics*[width=1.0 \textwidth]{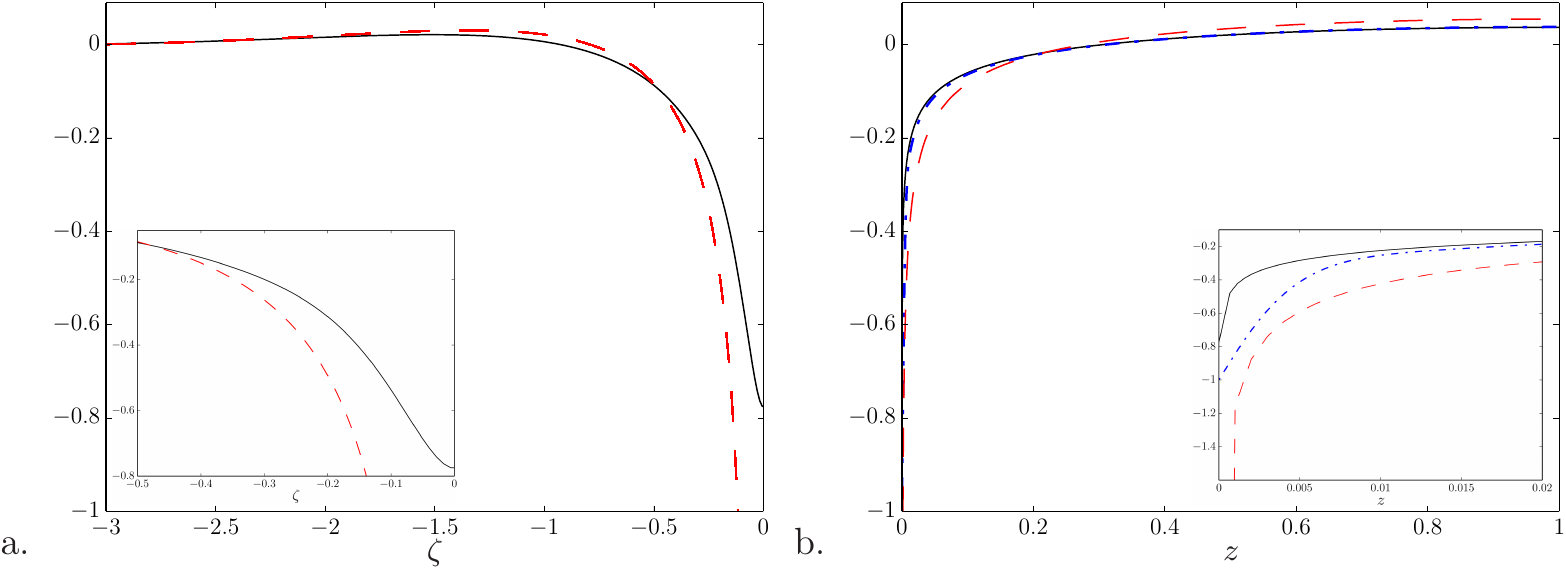}
}
\caption{(Colour online) For the case $E=10^{-5}$ ($\epsilon\approx 0.15$), $\ell=1$, profiles (ANS: black continuous; FS: red dashed; ENS: blue dot-dashed) of ($a$) $2\Psi_\tANS(\zeta,0)=\epsilon V_\tANS(\zeta,0)$ and $\epsilon V_\tFS(\zeta,0)$ versus $\zeta$, and ($b$) $\epsilon V_\tANS(0,z)$, $\epsilon V_\tENS(0,z)$ and $\epsilon V_\tFS(0,z)$ versus $z$. ({\itshape{Inset}}) Blow up ($a$) near $\zeta=0$, and ($b$) near $z=0$.}
\label{Psi_z=0}
\end{figure}

Finally in figure~\ref{ANS-ENS} we compare the ANS results $\Psi_\tANS$, $V_\tANS$ with the corresponding ENS-quantities $V_\tENS$ and $\Psi_\tENS$, defined by
\bse
\label{sl-var-ENS}
\begin{align}
\epsilon{\bar \omega}(\ell)\,V_\tENS(\zeta,z)\,=\,&\omega_\tENS(r,z)\,-\,{\bar \omega}(r)\,,\\
\ell^2{\bar \psi}(\ell)\,\Psi_\tENS(\zeta,z)(r,z)\,=\,&\psi_\tENS(r,z)\,-\,{\bar \psi}(r)(z-1)
\end{align}
\ese
(see (\ref{sl-var}$a$,$b$) and recall that ${\bar \psi}(r)=-\,\tfrac12\sigma r^2 {\bar \omega}(r)$,  (\ref{sl-var}$c$)), for the case $\ell=1$ and again $E=10^{-5}$. To extract the best possible approximation for $\Psi_\tENS$ from $\psi_\tENS$, we use $\sigma=1+(3/4)E^{1/2}$ (see (\ref{simple-spin-down}$c$)). The agreement is good and consistent with the $2\%$ estimated error from the neglect of terms order $O(\epsilon^2)$ neglected in the ANS-method. This success of the hybrid asymptotic-numerical method permits us to obtain results for small $E$ and overcomes the obstacle faced by the pure asymptotic method which requires $E$ to be essentially infinitesimally small.

An obvious test that addresses the question ``How useful is the zeroth order asymptotic approximation that gives $V_\tFS$?'' is to compare its lower $z=0$ and end $\zeta=0$ boundary values with $V_\tANS$ for a particular value of $E$. We do that in figures~\ref{Fourier-similarity}($a$,$b$) for $E=10^{-5}$ ($\epsilon\approx 0.1468$) and $\ell=1$, as in figures~\ref{ANS-fluct} and \ref{ANS-ENS}. More precisely we plot $2\Psi_\tANS(\zeta,0)=\epsilon V_\tANS(\zeta,0)$ in figure~\ref{Psi_z=0}($a$), where the $\epsilon$-scaling of $V$ is suggested by the corner boundary value $2\Psi_\tANS(0,0)=-1$ imposed by the end wall ($\zeta=0$) boundary condition (\ref {sl-bc}$a$)$_1$. Evidently this terminal value is approached but not reached by the ANS results and reflects the difficulties encountered by the numerics close to the corner singularity. The Fourier series solution $\epsilon V_\tFS(\zeta,0)$ plotted provides a valid approximation to $\epsilon V_\tANS(\zeta,0)$ provided that  $2\Psi_\tFS(\zeta,0)$ is small. Reasonable agreement is visible for $\zeta \lessapprox -0.4$ and the trend is acceptable for $\zeta \lessapprox  -0.15 \approx -\epsilon$, where $\epsilon V_\tFS(\zeta,0)\approx 3\epsilon/(\pi\zeta)$ (see (\ref{Y-Z-asym}$c$)). The $\zeta$-distance $\epsilon\approx 0.15$ identifies the lateral extent of the $E^{1/2}\times E^{1/2}$ corner region, inside which neither our FS or ANS results are valid. Similar considerations apply to the $\zeta=0$ boundary plots on figure~\ref{Psi_z=0}($b$), where the $z$-extent of the corner region is $\epsilon^3=E^{1/2}\approx 0.003$. With the consequent provisos, the comparisons of FS or ANS results reveal a similar level of accuracy. Once outside the corner region, the agreement of the ANS and ENS results, also portrayed on figure~\ref{Psi_z=0}($b$), is striking and provides strong evidence that ANS is a good approximation to the full ENS. 

\begin{figure}
\centerline{}
\vskip 3mm
\centerline{
\includegraphics*[width=1 \textwidth]{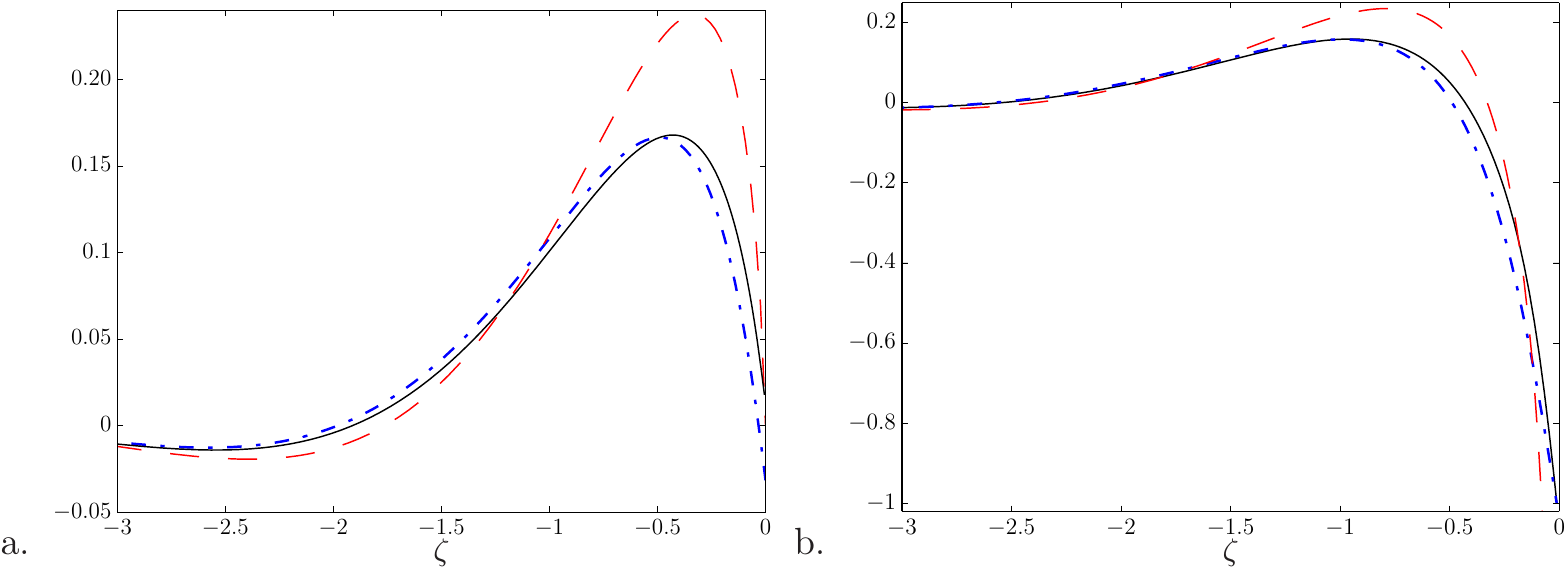}
}
\caption{(Colour online) As in figure~\ref{Psi_z=0}, profiles of the scaled $z$-mean values ${\bar V}=\epsilon^{-1}\langle V\rangle$,\quad ($a$)  ${\bar V}_{\!\!\tANS}(\zeta)$, ${\bar V}_{\!\!\tENS}(\zeta)$ and ${\bar V}_{\!\!\tFS}(\zeta)$, and ($b$) their derivatives $\dR {\bar V}/\dR\zeta$.}
\label{fig_6}
\end{figure}

Even sharper comparisons are made in figure~\ref{fig_6}($a$), again for $E=10^{-5}$ ($\epsilon\approx 0.1468$) and $\ell=1$, where the scaled $z$-average ${\bar V}=\epsilon^{-1}\langle V \rangle(\zeta)$ is plotted; a magnification by $\epsilon^{-2}=E^{-1/3}$ of the related $\epsilon V(\zeta,0)$ plotted in figure~\ref{Psi_z=0}($a$). The comparisons are of the ANS-solution ${\bar V}_{\!\!\tANS}(\zeta)=\epsilon^{-1}\langle V_{\tANS} \rangle$, the FS-solution ${\bar V}_{\!\!\tFS}(\zeta)=\int_{-\infty}^\zeta \Theta_\tFS(\zeta,0)\,\dR \zeta$ determined by (\ref{int-Xi}) and (\ref{sl-V-mean}$b$) and the ENS-solution ${\bar V}_{\!\!\tENS}(\zeta)$ which together with its derivative  $\dR{\bar V}_{\!\!\tENS}/\dR \zeta$ are defined just as in (\ref{sl-var-ENS}$a$,$b$) by
\bse
\label{sl-var-ENS-mean}
\begin{align}
E^{1/3}{\bar \omega}(\ell)\,{\bar V}_{\!\!\tENS}(\zeta)\,=\,&{\bar \omega}_\tENS(r)\,-\,{\bar \omega}(r)\,,\\
{\bar \omega}(\ell)\,\od{{\bar V}_{\!\!\tENS}}{\zeta}(\zeta)\,=\,&\od{{\bar \omega}_\tENS}{r}(r)\,-\,\od{{\bar \omega}}{r}(r)\,,
\end{align}
\ese
where ${\bar \omega}_\tENS(r)$ is defined by (\ref{omega-ENS}). Here, ${\bar \omega}_\tENS(r)$ and ${\bar \omega}(r)$ in (\ref{sl-var-ENS-mean}$a$), for our case $E=10^{-5}$, are plotted in figure~\ref{figk}($d$). We plot the derivatives $\dR {\bar V}_{\!\!\tANS}/\dR\zeta(\zeta)$, $\dR{\bar V}_{\!\!\tFS}/\dR\zeta(\zeta)=\Theta_\tFS(\zeta,0)$ and $\dR {\bar V}_{\!\!\tENS}/\dR\zeta(\zeta)$ in figure~\ref{fig_6}($b$).

The lowest order Fourier solutions ${\bar V}_{\!\!\tFS}$ and $\dR{\bar V}_{\!\!\tFS}/\dR\zeta$ are independent of $E$ and the reason for the choice of scaling in figures~\ref{fig_6}($a$,$b$). Very good agreement is obtained between the ANS and ENS results just as in figure~\ref{Psi_z=0}($b$). By contrast the Fourier series prediction is acceptable for $\zeta \lessapprox \,-1.5$ but is only fair for  $-1.5 \lessapprox \zeta <0$. This highlights the fact that ${\bar V}$ is a blow up by a factor $\epsilon^{-1}$ of the small quantity $\langle V\rangle$. This magnifies any errors and in part explains why it is difficult to obtain numerically accurate results from the asymptotics.

\begin{figure}
\centerline{}
\vskip 5mm
\centerline{
\includegraphics*[width=0.5 \textwidth]{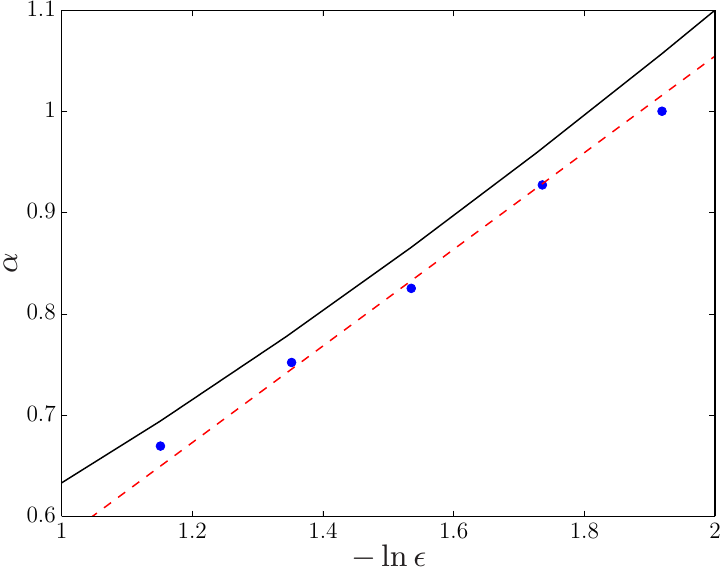}
}
\caption{(Colour online) Profile of $\alpha_\tANS$ versus $-\ln \epsilon =-(1/6)\ln E$ (black continuous curve). The blue bullets identify the values of $\alpha_\tENS$ ($\ell=1$) for $E=10^{-3}$, $3\times 10^{-4}$, $E=10^{-4}$, $3\times 10^{-5}$ $E=10^{-5}$ respectively (left to right). The best fit power law $\alpha\,=\,-\,(1/24\pi)\ln \epsilon\,+\,\mbox{const.}$ (see (\ref{alpha-pl})) is the red dashed line.}
\label{alpha-E}
\end{figure}

Our main objective is to determine $\alpha=\dR {\bar \omega}/\dR r(\ell)/{\bar \omega}(\ell)$ (see (\ref{kalpha}$c$)). We compare the ANS and ENS values 
\bme
\label{alpha-ANS-ENS}
\be
\alpha_\tANS\,=\,-\,\od{{\bar V}_{\!\!\tANS}}{\zeta}(0)\qquad\quad\mbox{and}\qquad \quad\alpha_\tENS\,=\,-\,\od{{\bar V}_{\!\!\tENS}}{\zeta}(0)\quad (\ell\,=\,1)
\ee
for various values of $E$ on figure~\ref{alpha-E}. The remarkably good agreement of $\alpha_\tANS$ and $\alpha_\tENS$ in figure~\ref{alpha-E} appears to reflect the apparently fortuitous agreement of these derivatives at $\zeta=0$ evident on figure~\ref{fig_6}($b$) for the particular case $E=10^{-5}$. For neighbouring values of $\zeta$ the agreement is not quite so good. Nevertheless, the generally favourable agreement of the ANS and ENS results illustrated by figures~\ref{fig_6}($a$,$b$)~and~\ref{alpha-E} vindicate our use of the hybrid asymptotical-numerical method-ANS. The corresponding true asymptotic value $\alpha_\tFS$ given by (\ref{sl-ln-E}) can only be extracted though the limiting procedure described thereabouts. In fact, our derivation of (\ref{sl-ln-E}) assumes that $\alpha_\tFS$ is determined, correct to the leading logarithmic order (but not the additional $O(1)$ constant part), by
\be
\se
-\,\od{{\bar V}_{\!\!\tFS}}{\zeta}(\zeta)\,\approx\,-\,\dfrac{3}{\pi}\ln(-\zeta)\,,
\ee
\eme
evaluated at $\zeta=\mbox{const.}\times\epsilon$ for any $O(1)$ constant. However, consideration of the $E=10^{-5}$ case illustrated in figure~\ref{fig_6}($b$) exposes the folly of this expectation at finite $E$. In the vicinity of $\zeta=-\epsilon\doteqdot -0.15$, the value of $\dR{\bar V}_{\!\!\tFS}/\dR\zeta$ varies rapidly and the choice of different $O(1)$ constants will determine enormously different values of $\alpha_\tFS$. Nevertheless as a somewhat futile gesture, we identified in figure~\ref{alpha-E} the power law 
\be
\label{alpha-pl}
\alpha\,=\,-\,(4\pi)^{-1}\ln E\,+\,\mbox{const.} 
\ee
that appeared to fit the ANS and ENS results. This differs by a factor $1/2$ from the asymptotic value $\alpha_\tFS=-(2\pi)^{-1}\ln E$ predicted by  (\ref{sl-ln-E}). It suggests, of course, that the power law (\ref{alpha-pl}) is an illusion representing a snapshot over a very limited range of $\epsilon$ which is not small anyway! One therefore concludes that $\alpha_\tFS$ is the true asymptotic limit but can only be reached by numerical values of $\epsilon$ that are essentially infinitesimals. Ultimately all our difficulties can be traced, as we remarked at the beginning of this subsection, to the singularity at $(\zeta,z)=(0,0)$ that is only properly resolved by consideration of the $E^{1/2}\times E^{1/2}$ corner region in its vicinity.

\section{Discussion\label{Discussion}}

As we explained at the outset, our investigation was motivated by the need to understand the role of artificial lateral boundaries in the modelling of essentially unbounded atmospheric flows. Our investigation of the spin-down problem suggests that the introduction of a stress-free outer boundary is far from adequate to obtain realism because motion is seriously affected by the presence of any impermeable boundary. The point here is that the classic spin-down mode (\ref{simple-spin-down}$a$) exhibits a radial QG-outflow $u^\star\propto r^\star$. So any boundary condition that prevents that natural outflow will have a profound effect as our analysis has shown. Indeed spin-down is a feature of vortex line shortening (equivalently fluid columns broadening). However the forced return flow in the $E^{1/3}$--side-wall shear-layers has exactly the opposite effect causing a slight slow down of the spin-down process which is more pronounced towards the outer boundary. In summary good modelling of the unbounded system needs careful consideration of the fluid flux side-wall boundary condition upon which some results will be sensitive.

We have studied the entire transient process numerically. Our overall DNS-results are reported in \S\ref{DNS}, while the final asymptotic form of the solution is provided by the ENS-results reported in \S\ref{ENS}. Our asymptotic confirmation of those results is only partially achieved. For though the small-$E$ mainstream analytic solution is 
\be
\label{ms-as-sol}
\bar{\breve \omega}\,=\,\kappa \,\dfrac{2{\mathring \alpha}}{k^2\,-\,2{\mathring \alpha}\,-\,{\mathring \alpha}^2}\,\dfrac{\IR_1(k r/\ell)}{(r/\ell)\,\IR_1(k)}\,\exp\bigl[\bigl(-\sigma\ell^{2} E^{-1/2}+\,k^2\bigr)\tau\bigr] \qquad \mbox{as}\quad \tau\to\infty 
\ee
(see (\ref{ms-lt}), (\ref{sol-tau}), (\ref{Bessel-sol}) and (\ref{left-right-end-values}), the values of $k$ and ${\mathring \alpha}(k)\equiv k\IR_2(k)\big/\IR_1(k)=\ell\alpha$ (see (\ref{kalpha}$b$,$d$)) characterising (\ref{ms-as-sol}) are only fixed after $\alpha$ is determined by the solution of the $E^{1/3}$-layer problem (see \S\ref{side-wall}), the asymptotic solution of which is possible (see (\ref{alpha-pl})) but only useful at extremely small $E$. For the moderately small $E$ used in the DNS and ENS, it is necessary to solve the shear-layer equations numerically. The \S\ref{side-wall-hybrid} results determine $\alpha=\alpha(E)$ dependant on $E$ alone and hence complete our hybrid asymptotic-numerical solution. 

In contrast to (\ref{ms-as-sol}), the solution (\ref{QG-me-GH}$a$,$b$) (cf., (\ref{Bessel-sol}) and (\ref{Bessel-A0})) for the rigid outer boundary case considered by \cite{GH63} determines
\be
\label{ms-as-sol-GH}
\bar{\breve \omega}\,=\,2 \kappa\,\dfrac{\JR_1(j_1 r/\ell)}{(j_1r/\ell)\,\JR_2(j_1)}\,\exp
\bigl[\bigl(-\sigma\ell^{2} E^{-1/2}-\,j_1^2\bigr)\tau\bigr] \qquad \mbox{as}\quad \tau\to\infty\,,
\ee
where $j_1=3.8317\cdots$ is a constant, unlike $k$ which is a function of $\ell$ and $E$ being the solution of ${\mathring \alpha}(k)=\ell\alpha(E)$. The derivation of (\ref{ms-as-sol-GH}) is similar to (\ref{ms-as-sol}), particularly (\ref{Bessel-sol})--(\ref{left-right-end-values}), where now we need $\int_0^{j_1} \rho \bigl[\JR_1(\rho)\bigr]^2\,\dR\rho= \tfrac12 j_1^2\bigl[\JR_2(j_1)\bigr]^2$ \citep[see][\S7.1.4, eq.~(48)]{EMOT53}. Furthermore, comparison of the spatial and temporal differences between the solutions (\ref{ms-as-sol}) and (\ref{ms-as-sol-GH}) is striking and reinforces remarks made in \S\ref{GH63-problem}.

Though the $\tau\to\infty$ limit taken in (\ref{ms-as-sol}) and (\ref{ms-as-sol-GH}) highlights the differences between the stress-free and rigid outer boundary cases, \cite{GH63} never derived (\ref{ms-as-sol-GH}), because by that late time (even long on the long lateral diffusion time-scale $\ell^2E^{-1}/\Omega$) the solution, for all intents and purposes, is negligible. They were concerned with the solution on the shorter spin-down time-scale $E^{-1/2}/\Omega$ on which the solution is still finite. Despite our analysis in \S\ref{QG-trans} of the evolution of the scaled QG-azimuthal angular velocity $\bar{\mathring \omega}$ being cast on the $\ell^2E^{-1}/\Omega$ time-scale, it also pertains to the relatively short time $E^{-1/2}/\Omega$, provided it is long compared to the time $E^{-1/3}/\Omega$ needed to establish the $E^{1/3}$-layer. It is therefore illuminating to consider the force balances controlling the unscaled angular velocity $\bar{\breve \omega}$ on the shorter spin-down time-scale $E^{-1/2}/\Omega$ by expressing (\ref{QG-eq-new}) in its more primitive form
\be
\label{QG-eq-new-sd-time}
E^{-1/2}\pd{\bar{\breve \omega}}{t}\,+\,\sigma\bar{\breve \omega}\,=\,\dfrac{E^{1/2}}{r^3}\pd{\,}{r}\biggl(r^3\pd{\bar{\breve \omega}}r\biggr).
\ee
From the balance of the Ekman suction term $\sigma\bar{\breve \omega}$ (bottom friction) on the left-hand side and the lateral diffusion term on the right-hand side, we identify the classical $E^{1/4}$ \cite{S57} layer. Indeed Barcilon solved (\ref{QG-eq-new-sd-time}), expressed in boundary layer form \citep[see][eqs.~(17), (18)]{B68}, subject to Stewartson-like boundary conditions for which there is a steady $E^{1/4}$-layer in the limit $E^{-1/2}t\to\infty$. \cite{GH63} also refer to their shear-layer solution on the $E^{-1/2}/\Omega$ time-scale as an  $E^{1/4}$-layer. Their primary balance, however, is between the terms on the left-hand side of (\ref{QG-eq-new-sd-time}) that describe spin-down without identifying any length scale. That fast spin-down is modulated on the slow time $\tau$ (see (\ref{two-timing})), a partition often referred to as two-timing. Then, as the scaled equation (\ref{QG-eq-new}) shows, that modulation diffuses on the lateral diffusion length $\delta^\star=H(Et)^{1/2}$, which just happens to equal $O(HE^{1/4})$ on the spin-down time $t=E^{-1/2}$. For $t\gg E^{-1/2}$ the layer slowly thickens until the solution attains the asymptotic form (\ref{ms-as-sol}). Be that as it may, we reiterate our remarks of \S\ref{Outline} that the \S\ref{QG-series} power series solution is valid on the spin-down time $E^{-1/2}/\Omega$ and provides our analogue of the rigid boundary solution \citep[][eq.~(6.3)]{GH63}.

The cornerstone of our study is the investigation in \S\ref{side-wall} of the $E^{1/3}$-layer, whose main purpose is the determination of  $\alpha$ needed for the QG-mainstream boundary condition $\dR{\bar\omega}/\dR r(\ell)=\alpha{\bar\omega}(\ell)$ (see (\ref{kalpha}$c$)). Nevertheless, amongst the results reported there, the power series solution (\ref{Xi-Gamma-sim}$a$,$b$) of the zeroth order shear-layer problem in \S\ref{0-ps} is of interest in its own right, as it provides an alternative to the usual Fourier series expansion  (\ref{sl-sol-0}$a$,$b$). Being constructed as the sum of powers of $z$, each multiplied by functions of the similarity variable $\Phi=(Ez)^{-1/3}(r-\ell)$ (see (\ref{Xi-Gamma-sim}), (\ref{Xi-Gamma-N}) and (\ref{Xi-Gamma-sim}$c$)), it is especially accurate in the vicinity of the singularity at $(r,z)=(\ell,0)$, where the Fourier series fails or at any rate needs a very large number of terms to converge. Indeed only a very few terms of the power series are needed to give good results throughout most of the shear-layer (see figure~\ref{Fourier-similarity}). It should , however, be stressed that the power series solution is only possible because we have the explicit Fourier forms (\ref{Xi-Gamma-0}$a$--$e$) at $r=\ell$ to build on.

Though we relegated our extension of the \cite{GH63} transient solution of the infinite plane layer problem to appendix~\ref{infinite-layer}, our new results are both intriguing and useful. The fact that the large time solution (\ref{inf-layer-sol}$a$) involves the factor $\kappa= 1+O(E^{1/2})\not=1$ (see (\ref{kappa-mu}$a$)) surprised us, so prompting the question ``Where does the $O(E^{1/2})$ contribution to $\kappa$ come from?''. The possibility, that it originates from the early time behaviour before the Ekman layer has formed, can be dismissed because the Laplace transform solution of appendix~\ref{infinite-layer} only appeals to a full developed Ekman layer (see (\ref{LT-sol}$b$)). Instead the small $O(E^{1/2})$ contribution to $\kappa$ must be a consequence of inertial wave radiation. Certainly inertial waves are visible in the full numerical solution but diminish in size relative to the QG-part as time proceeds. It should be noted that the positive sign of $\kappa-1\approx E^{1/2}/4$ indicates that the QG-flow has decayed less than expected consistent with energy being transferred into inertial waves rather than taken from the QG-flow. 

Significantly, the introduction of the factor $\kappa$ early on in (\ref{sol-tau}) has led to remarkable agreement between the transient amplitudes of the DNS-results and the numerical solution of the QG-equation (\ref{QG-eq-new}) displayed in figure~\ref{transient}, which to graph plotting accuracy are virtually indistinguishable except within the $E^{1/3}$-layer. Likewise, the factor $\mu= 1-\tfrac12 E^{1/2}$ (see (\ref{kappa-mu}$b$)) has proved useful in the definitions (\ref{omega-ENS}) and (\ref{omega-DNS}) of ${\bar\omega}_{\tENS}(r)=\mu^{-1}\langle \omega_{\tENS}\rangle(r)$ and ${\bar{\breve \omega}}_{\tDNS}(r,t)=\mu^{-1}\langle {\breve \omega}_{\tDNS}\rangle(r,t)$ respectively, again leading to improvements in the comparisons with our numerics and asymptotics in all cases.

\section*{Acknowledgements}

This project was partially supported by the French National Center for Scientific Research (CNRS) under the interdisciplinary grant Inphyniti/MI/CNRS. The numerical simulations were performed using HPC resources from GENCI-IDRIS (Grant 2015-100584 and 2016-100610).


\appendix

\section{Final decay of QG-motion in an unbounded plane layer\label{infinite-layer}}

We extract the salient details of solution of the spin-up problem considered by \cite{GH63} for two parallel coaxial infinite disks in their section 3 and extend them as needed for the infinite plane layer solution of our initial value spin-down problem (\ref{mom-vort}-\ref{bc}). Like \cite{GH63} we assume the similarity form
\bse
\label{LT}
\be
[\,{\breve u}\,,\,{\breve v}\,] = r[\,{\mathfrak u}\,,\,{\mathfrak v}\,](z,t)
\ee
and solve by the Laplace transform
\be
 [\,\hat{\mathfrak u}\,,\,\hat{\mathfrak v}\,](z,p)\,=\,\int_0^\infty[\,{\mathfrak u}\,,\,{\mathfrak v}\,](z,t)\,\exp(-pt)\,\dR t
\ee
\ese
method. We define
\bme
\label{LT-sol}
\se
\be
E^{1/2}m_\pm\,=\,(p\pm 2\iR)^{1/2}\,=\,(1\pm\iR)\bigl(1\mp\tfrac12\iR p\bigr)^{1/2}
\ee
\citep[cf.][eq.~(3.6), in which $(m_1,m_2)\mapsto(m_+,m_-)$]{GH63}  so that, provided that $|m_\pm|=O\bigl(E^{-1/2}\bigr)\gg 1$, the solution may be written
\be
\hat{\mathfrak v}\,\mp\,\iR\,\hat{\mathfrak u}\,=\,2D^{-1}m_\pm (m_\mp-1)\bigl(1-\exp(-m_\pm z)\bigr), 
\ee
where
\begin{align}
D(p)\,=\,E(m_+^3(m_--1)\,+\,m_-^3(m_+-1))\,=\,&\,2pm_+m_-\,-\,E\bigl(m_+^3+m_-^3\bigr)\qquad\\
=\,&\,2M(p)\,+\,pN(p)
\end{align}
\be
\de
M(p)\,=\,\iR(m_--m_+)\,,\qquad\qquad\quad N(p)\,=\,m_+(m_--1)\,+\,m_-(m_+-1)
\ee
\eme
\citep[cf.][particularly eq.~(3.11)]{GH63}.

The $E^{-1/2}z\to\infty$ limit of (\ref{LT-sol}$b$) determines the mainstream values
\bme
\label{LT-mainstream}
\be
\hat{\mathfrak u}\,\to\, \hat{\bar{\mathfrak u}}\,=\,M/D\,, \qquad\qquad
\hat{\mathfrak v}\,\to\, \hat{\bar{\mathfrak v}}\,=\,N/D\,,\qquad\qquad
\ee
\eme
while the $z$-average (see (\ref{MS-flux}$b$)) gives
\bme
\label{LT-average}
\be
\te
\bigl\langle\hat{\mathfrak u}\bigr\rangle\,=\,0\,,\qquad\qquad
\bigl\langle\hat{\mathfrak v}\bigr\rangle\,=\,L/D\,,\qquad\qquad
L(p)=\,2(m_+-1)(m_--1)\,,
\ee
\eme

The QG-flow for $t=O\bigl(E^{-1/2}\bigr)\gg 1$ is determined by the residue at the pole $p=p_0$, where $D(p)=0$, of the inverse Laplace transform. It yields
\bse
\label{inf-layer-sol}
\be
\bigl[\,\bar{\mathfrak u}\,,\,\bar{\mathfrak v}\,\bigr]\,=\,[M(p_0),\,N(p_0)]\,\exp(p_0t)\big/D^{\,\prime}(p_0)\,
=\,\kappa\,\bigl[\,-\,\tfrac12 p_0\,, \,1\,\bigr]\exp(p_0t)\,,
\ee
where 
\be
\kappa\,=\,N(p_0)\big/D^{\,\prime}(p_0)\,, \qquad\qquad\qquad \bigl(\,D^\prime=\dR D/\dR p\,\bigr)\,.
\ee
\ese
The $z$-average $\langle{\mathfrak v}\rangle$ is related the mainstream value $\bar{\mathfrak v}$ by
\bme
\label{inf-layer-sol-av}
\be
\langle{\mathfrak v}\rangle\,=\,\mu\,\bar{\mathfrak v}\,,\qquad\qquad\mbox{where}\qquad \qquad \mu\,=\,L(p_0)/N(p_0)\,.
\ee
\eme
Assuming that $p=O\bigl(E^{1/2}\bigr)$, we obtain
\bme
\label{NandM}
\be
\left.\begin{array}{rl}
EN(p)\,=&\,4\,-\,2 E^{1/2}\,+\,O(E)\,, \\[0.2em]
EL(p)\,=&\,4\,-\,4 E^{1/2}\,+\,O(E)\,,
\end{array}\right\}
 \qquad\quad
ED^{\,\prime}(p)\,=\,4\,-\,3 E^{1/2}\,+\,O(E)\,.
\ee
\eme

We set $p_0=-E^{1/2}\sigma$ (real $\sigma$) and substitute into $p=E\bigl(m_+^3+m_-^3\bigr)\big/(2m_+m_-)$ (see (\ref{LT-sol}$c$)) to obtain
\bse
\label{p-sigma}
\begin{align}
\sigma\,=\,-\,\Re\biggl\{\dfrac{E^{1/2}m_-^2}{m_+}\biggr\}\,=\,&\,\Re
\biggl\{\dfrac{(1+\iR)\bigl(1-\iR\tfrac12 \sigma E^{1/2}\bigr)}
{\bigl(1+\iR\tfrac12 \sigma E^{1/2}\bigr)^{1/2}}\biggr\}
\intertext{with the solution expansion}
\sigma\,=\,&\,1\,+\,\tfrac34 E^{1/2}\,+\,O(E)\,.
\end{align}
\ese
In addition (\ref{inf-layer-sol}$b$), (\ref{inf-layer-sol-av}$b$) and (\ref{NandM}$a$,$b$) determine
\bme
\label{kappa-mu}
\be
\kappa\,=\,1\,+\,\tfrac14 E^{1/2}\,+\,O(E)\,, \qquad\qquad
\mu\,=\,1\,-\,\tfrac12 E^{1/2}\,+\,O(E)\,.
\ee
\eme

\vskip 2cm

\section{The small $\tau$ power series solution\label{QG-pow-ser}}

To solve (\ref{QG-eq-new})-(\ref{QG-bc}) for $\tau\ll 1$, we make the preliminary change of variables
\bme
\label{F-G}
\be
{\bar{\mathring \omega}}\,=\,1\,+\,f(\zeta,\tau)F(\zeta)\,+\,g(\zeta,\tau) G(\zeta), \qquad\quad G^{\,\prime}(\zeta)\,=\,-\,F(\zeta)\,,
\ee
where
\be
F(\zeta)\,=\,\bigl(2\big/\sqrt{\pi}\bigr)\exp(\,-\zeta^2), \qquad\qquad
G(\zeta)\,=\erfc\zeta\,=\,\int_\zeta^\infty F(\tilde{\zeta})\, \dR\tilde{\zeta}\,.
\ee
\eme
It satisfies (\ref{QG-eq-new}) when
\bse
\label{sim-form-eq-fg}
\begin{align}
\pd{^2f}{\zeta^2}\,-\,2\zeta\pd{f}{\zeta}\,-\,2g\,-\,4\tau\pd{f}{\tau}\,-\,2\pd{g}{\zeta}\,=\,&\,\dfrac{6\tau^{1/2}}{1-2\tau^{1/2}\zeta}\biggl(\pd{f}{\zeta}\,-\,2\zeta f\,-\,g\biggr),\\
\pd{^2g}{\zeta^2}\,+\,2\zeta\pd{g}{\zeta}\,-\,4\tau\pd{g}{\tau}\,=\,&\,\dfrac{6\tau^{1/2}}{1-2\tau^{1/2}\zeta}\,\pd{g}{\zeta}\,,
\end{align}
where
\be
\dfrac{1}{1-2\tau^{1/2}\zeta}\,=\, 1+\sum_{n=1}^\infty (4\tau)^{n/2}\zeta^n\,.
\ee
\ese
The boundary condition (\ref{QG-bc}) at $r=\ell$ is satisfied when
\be
\dfrac{2}{\sqrt\pi}\biggl(\pd{f}{\zeta}\,-\,g\biggr)+\,\pd{g}{\zeta}\,=
-\,\mathring\alpha\sqrt{4\tau}\biggl(1\,+\,\dfrac{2}{\sqrt\pi}f\,+\,g\biggr) \qquad\qquad \mbox{at} \qquad\zeta=0\,.
\label{sim-form-bc-fg}
\ee

We express each $\varpi_n(\zeta)$ in the solution (\ref{sim-series}) for ${\bar{\mathring \omega}}$ in the form
\be
\label{sim-series-n}
\varpi_n\,=\,f_n(\zeta)F(\zeta)\,+\,g_n(\zeta) G(\zeta)\,,
\ee
so that
\be
\label{sim-series-fg}
{\bar{\mathring \omega}}\,=\,1\,+\,{\mathring \alpha}\sum_{n=1}^\infty (4\tau)^{n/2} \bigl[f_n(\zeta)F(\zeta)\,+\,g_n(\zeta) G(\zeta)\bigr],
\ee
and seek polynomial solutions
\bse
\label{sim-series-nm}
\begin{align}
f_{1}(\zeta)\,=&\,f_{10}\,,&    g_{1}(\zeta)\,=&\,g_{11}\zeta\,,\\
f_{2}(\zeta)\,=&\,f_{21}\zeta\,,&   g_{2}(\zeta)\,=&\,g_{20}+g_{22}\zeta^2\,,\\
f_{3}(\zeta)\,=&\,f_{30}+y_{32}\zeta^2\,,&  g_{3}(\zeta)\,=&\,g_{31}\zeta+g_{33}\zeta^3\,,\\
f_{4}(\zeta)\,=&\,f_{41}\zeta+f_{43}\zeta^3\,,&  g_{4}(\zeta)\,=&\,g_{40}+g_{42}\zeta^2+g_{44}\zeta^4\,,
\end{align}
\ese
in which $f_{nm}$ and  $g_{nm}$ are constants. The polynomials for $n>4$ can be expressed in the obvious way. They determine the end point ($r=\ell$) value
\be
\label{sim-series-ell_A}
{\bar{\mathring \omega}}(\ell,\tau)\,=\,1\,+\,\dfrac{\mathring \alpha}{\sqrt{\pi \tau}}\,\sum_{q=1}^\infty f_{(2q-1)\,0}\,(4\tau)^q\,+\,{\mathring \alpha}\sum_{q=1}^\infty g_{(2q)\,0}\,(4\tau)^q
\ee
given by (\ref{sim-series-ell}) for $n\le 4$ on employing the results below.

Substitution of (\ref{sim-series-n}) with (\ref{sim-series-nm}) into the equations (\ref{sim-form-eq-fg}) and boundary condition (\ref{sim-form-bc-fg}), and then equating coefficients at each respective power of $\tau$ leads to 

\bigskip
\noindent At $O\bigl(\tau^{1/2}\bigr)$, $\quad n=1$:
\be
\label{sim-form-ser-fg-1}
f_{10}\,=\,\tfrac12\,,\qquad\qquad\qquad\qquad\quad g_{11}\,=\,-1\,.\qquad\qquad\qquad\,\,\,\,
\ee
\noindent  At $O\bigl(\tau\bigr)$, $\quad n=2$:
\bse
\label{sim-form-ser-fg-2}
\begin{align}
f_{21}=\,&\dfrac{1}{4}\biggl(\dfrac32-{\mathring \alpha}\biggr), & 
g_{20}=\,&\dfrac{1}{4}\biggl(\dfrac32+{\mathring \alpha}\biggr),\qquad\qquad\\  
&&g_{22}=\,&-\dfrac{1}{2}\biggl(\dfrac32-{\mathring \alpha}\biggr).
\end{align}
\ese
\noindent  At $O\bigl(\tau^{3/2}\bigr)$, $\quad n=3$:
\bse
\label{sim-form-ser-fg-3}
\begin{align}
f_{30}=&\dfrac{1}{16}\biggl(\dfrac52+4{\mathring \alpha}+\dfrac43 {\mathring \alpha}^2\biggr),&
g_{31}=&-\dfrac{\mathring \alpha}{4}\biggl(\dfrac32+{\mathring \alpha}\biggr),\\
f_{32}=&\dfrac{1}{4}\biggl(\dfrac{5}{2}-\dfrac{{\mathring \alpha}}{2}+\dfrac{{\mathring \alpha}^2}{3}\biggr),&
g_{33}=&-\dfrac{1}{6}\biggl(\dfrac{15}{2}-\dfrac32{\mathring \alpha}+{\mathring \alpha}^2\biggr).
\end{align}
\ese
\noindent At $O\bigl(\tau^2\bigr)$, $\quad n=4$:
\bse
\label{sim-form-ser-fg-4}
\begin{align}
f_{41}=\,&\dfrac{1}{64}\biggl(\dfrac{15}{4}+{2}{\mathring \alpha}-7{\mathring \alpha}^2-\dfrac{10}{3}{\mathring \alpha}^3\biggr),&
g_{40}=\,&\dfrac{3}{64}\biggl(\dfrac{5}{4}+4{\mathring \alpha}+3{\mathring \alpha}^2+\dfrac{2}{3}{\mathring \alpha}^3\biggr), \\
f_{43}=\,&\dfrac{1}{64}\biggl(\dfrac{95}{2}-16{\mathring \alpha}+2{\mathring \alpha}^2-\dfrac{4}{3}{\mathring \alpha}^3\biggr),&
g_{42}=\,&\dfrac{3}{32}\biggl(\dfrac{5}{2}+2{\mathring \alpha}+2{\mathring \alpha}^2+\dfrac{4}{3}{\mathring \alpha}^3\biggr),\\
&&g_{44}=\,&\dfrac{1}{32}\biggl(\!-\dfrac{95}{2}+16{\mathring \alpha}-2{\mathring \alpha}^2+\dfrac{4}{3}{\mathring \alpha}^3\biggr).
\end{align}
\ese
Though their derivation is straightforward, that for each successive order $O\bigl(\tau^{n/2}\bigr)$ becomes increasingly tedious, which is why we have given explicit results for $n=1,\cdots,4$.

\vskip 2cm
\section{Power series for $\Psi$ and $V$ for $N\le 4$\label{similarity-4}}

We provide explicit series solutions for $\Psi=\Im\{\Xi\}$ and $V=\Re\{\Upsilon\}$, where $\Xi$ and $\Upsilon$ are defined by (\ref{Xi-Gamma-sim}$a$,$b$). We truncate the series (\ref{Xi-Gamma-N}$a$,$b$) for $\Xih_N(\Phi,z)$, $\Upsilonh_N(\Phi,z)$ that approximate $\Xih(\Phi,z)$, $\Upsilonh(\Phi,z)$ at $N=4$ to obtain
\bse
\label{Psi-V-origin-more}
\begin{align}
\Psi(\zeta,z)\,=\,&\,\Im\bigl\{\YS(\Phi)\bigr\}+\,\dfrac12(-1+z)\,
+\,\dfrac{\Gamma(4/3)\zeta(4/3)}{2\pi}z\zeta\,-\,\dfrac{\pi}{36}z\zeta^3\,\nonumber\\
&-\,\dfrac{\sqrt{3}\,\Gamma(7/3)\,\zeta(7/3)}{48\pi}z\zeta^4
\,+\,\dfrac{\Gamma(10/3)\,\zeta(10/3)}{48\pi}\,z\biggl[z^2\,-\,\dfrac{1}{210}\zeta^6\biggr]\zeta\,\nonumber\\
&\,-\,\dfrac{\pi^3}{2160}\,z\biggl[z^2\,-\,\dfrac{1}{2520}\zeta^6\biggr]\zeta^3\nonumber\\
&\,-\,\dfrac{\sqrt3\, \Gamma(13/3)\,\zeta(13/3)}{1152\pi}\,z\biggl[z^2\,-\,\dfrac{1}{6300}\zeta^{6}\biggr]\zeta^4\nonumber\\
&\,+\,\OR\bigl(z^{16/3}\bigr)\,,\\[0.5em]
V(\zeta,z)\,=\,&\,z^{-1/3}\Re\bigl\{\ZS(\Phi)\bigr\}-\,\dfrac{2^{1/3}}{\pi^{2/3}} \zeta(2/3)
\,-\,\tfrac12\zeta^2\,-\dfrac{\Gamma(4/3)\zeta(4/3)}{6\pi}\zeta^3\,+\,\dfrac{\pi}{360}\zeta^5\,\nonumber\\
&\,-\,\dfrac{\sqrt{3}\,\Gamma(7/3)\,\zeta(7/3)}{8\pi}\biggl[z^2\,-\,\dfrac{1}{90}\zeta^6\biggr]
-\,\dfrac{\Gamma(10/3)\,\zeta(10/3)}{48\pi}\biggl[z^2\,-\,\dfrac{1}{7560}\zeta^6\biggr]\zeta^3\nonumber\\
&\,+\,\dfrac{\pi^3}{7200}\biggl[z^2\,-\,\dfrac{1}{41580}\zeta^6\biggr]\zeta^5\nonumber\\
&\,-\,\dfrac{\sqrt3\, \Gamma(13/3)\,\zeta(13/3)}{384\pi}\biggl[z^4\,-\dfrac{1}{15}z^2\zeta^6\,+\,\dfrac{1}{1247400}\zeta^{12}\biggr]\nonumber\\
&\,+\,\OR\bigl(z^{5}\bigr)\,
\end{align}
\ese
for $\Phi=\OR(1)$.

\end{document}